\documentclass[11pt,a4paper,oneside]{article}
\usepackage{lineno,hyperref}

\usepackage{graphicx}              % to include figures
\usepackage{epsfig}
\usepackage{amsmath}               % great math stuff
\usepackage{amssymb}
\usepackage{epstopdf}
\usepackage{verbatim}
\usepackage{mathptmx}
\usepackage{mathalpha}
\usepackage{mathtools}
\usepackage[scale=0.80]{geometry}
\usepackage{graphicx, times}
\usepackage{amsfonts}              % for blackboard bold, etc
\usepackage{amsthm}                % better theorem environments\right 
\usepackage{multicol}
\usepackage{algorithm}
\usepackage{algorithmic}
\usepackage{varwidth}
\usepackage{parskip}
\usepackage{hyperref}
\usepackage{rotating}
\usepackage[numbers,sort&compress]{natbib}
\usepackage{multirow}
\usepackage{pdflscape}
\usepackage[numbers]{natbib}
\usepackage{caption}
\usepackage{xcolor}
\usepackage{color}
\usepackage{comment}
\usepackage{subcaption}
\newcommand*{\rom}[1]{\expandafter\@\romannumeral #1}
\newcommand{\bea}{\begin{eqnarray}}
	\newcommand{\eea}{\end{eqnarray}}
\newcommand{\bee}{\begin{eqnarray*}}
	\newcommand{\eee}{\end{eqnarray*}}

\begin{document}
\author{Khomesh R. Patle$^{1}$\footnote{khomeshpatle5@gmail.com}, G. P. Singh$^{1}$\footnote{gpsingh@mth.vnit.ac.in}, Romanshu Garg$^{1}$\footnote{romanshugarg18@gmail.com}
\vspace{.3cm}\\
${}^{1}$ Department of Mathematics,\\ Visvesvaraya National Institute of Technology, Nagpur, 440010, Maharashtra India.
\vspace{.3cm}
%\\${}^{2}$ Departments of Mathematics, \\ Jadavpur University, Kolkata 700032, West Bengal, India.
\date{}}
	
%${}^{2}$ Departments of Mathematics,\\
%Jadavpur University, Kolkata 700032, West Bengal, India.}
%\title{Accelerated expansion of the universe model with quadratic parametrization of Hubble parameter in $f(R,L_m)$ theory of gravity}
\title{Accelerated expansion of the universe model with parametrization of $q(z)$ in $f(R,L_m)$ theory of gravity}

\maketitle
\begin{abstract} \noindent
In this research work, we explore the late-time accelerated expansion of the universe within the framework of modified gravity, specifically $f(R, L_m)$ theory, by considering two non-linear models: $f(R, L_{m}) = \frac{R}{2}+(1+\eta R) L_{m}$ and $f(R, L_{m}) = \frac{R}{2}+ L_{m}^{\eta}$, where $\eta$ is a free parameter. Adopting a parametric form of the deceleration parameter $q(z)$, we derive a quadratic expression for the normalized Hubble parameter. 
%and incorporate it into the modified Friedmann equations.
By use of Bayesian statistical analysis with the $\chi^{2}$-minimization approach, we determine the median values of the model parameters for
%by fitting to
both the cosmic chronometer (CC) and the joint (CC+Pantheon) dataset. 
%The analysis confirms a transition from decelerated to accelerated expansion at $z_{t}=0.655$, with present-day values of $q_{0}$ consistent with late-time cosmic acceleration.
Furthermore, we examine the fundamental cosmological parameters: energy density, pressure, the equation of state (EoS) parameter, and energy conditions. The cosmographic parameters are thoroughly analyzed, and the present age of the universe is estimated based on this model.
%The EoS and kinematic parameters indicate a quintessence-like dark energy behaviour for both models and a slight deviation from the standard $\Lambda$CDM model, with Model(II) showing stronger acceleration and better agreement with dark energy-dominated dynamics. Overall, these results support the effectiveness of the chosen models in describing the current cosmological dynamics.
%Overall, both models are physically viable and effectively capture the essential features of the current cosmic acceleration.
\end{abstract}

\section{Introduction}\label{sec:1}
Astronomical research~\cite{1998AJ....116.1009R,1999ApJ...517..565P,2020A&A...641A...6P} has verified the accelerating expansion of the universe, marking a fascinating revelation about the observable cosmos. Subsequently, theoretical cosmologists developed an interest in designing models that illustrate the universe's accelerated expansion phase.
%In an effort to account for the universe's acceleration, numerous theoretical cosmologists have developed models that adjust Einstein’s field equations or explore alternative gravitational theories.
The discovery of cosmic acceleration has motivated many researchers either to modify Einstein’s field equations or investigate alternative theories of gravity, aiming to better understand the underlying physics driving this expansion.
%To explain the universe’s accelerated expansion, many researchers have proposed modifying Einstein’s field equations or exploring alternative gravity theories. These models aim to address the limitations of standard cosmology and provide a better understanding of cosmic acceleration.
The exact identity of the enigmatic energy causing the universe to expand at an accelerated rate is still unknown, leading most researchers to call it dark energy. A refined cosmological model has been suggested to explain the universe's accelerating expansion, drawing on various forms of dark energy that exhibit negative pressure and positive energy density. The dynamical cosmological term $\Lambda$ is broadly recognized as one of the leading contenders for dark energy~\cite{weinberg1989cosmological}. However, the $\Lambda$ term model struggles with some issues, including fine-tuning and coincidence problems~\cite{weinberg1989cosmological,di2021realm,carroll2001cosmological}. In recent decades, researchers have explored different strategies to tackle modern cosmological issues. Currently, the most commonly applied approach to solving modern cosmological problems is the modified gravity theory. To elucidate the mysteries of dark content, researchers have developed $f(R)$ gravity~\cite{buchdahl1970non} 
%a theoretical construct that generalizes the principles of General Relativity and provides new insights into the behavior of cosmic structures.
a generalized theoretical construct that builds upon the foundations of General Relativity, aiming to yield new perspectives on the formation and behavior of large-scale cosmic structures.
Concurrently with the advancement of knowledge, various modified theories have garnered significant attention and investigation~\cite{harko2011f,nojiri2011unified,nojiri2017modified,capozziello2011cosmography,capozziello2019extended,lalke2023late,kotambkar2017anisotropic,singh1997new,singh2025observational,varela2025cosmological,singh2022cosmic,hulke2020variable,singh2024generalized,bamba2010finite,garg2025cosmic,singh2018thermodynamical}, reflecting the dynamic nature of theoretical development in this field.
%,capozziello2023role
\vspace{0.1cm}\\
A novel theoretical approach was established by relating the Ricci scalar $R$ to the matter Lagrangian density $L_m$, giving rise to the $f(R, L_m)$ gravitational Lagrangian density that encapsulates the gravitational-matter interaction. As highlighted by Harko and Lobo~\cite{harko2010f}, this theory offers a more general and flexible framework for investigating the dynamic interplay between matter and geometry, facilitating the discovery of novel connections and relationships. Moreover, a fundamental shortcoming of this approach is its inability to reconcile with the equivalence principle. The precision of solar system observations has been shown to significantly restrict the parameter space of theoretical models that violate fundamental physical principles~\cite{Faraoni2004pi,zhang2007behavior,bertolami2008general,Rahaman2009solar}. The $f(R, L_m)$framework is a theoretically comprehensive and empirically well-supported gravitational theory that has been developed within the mathematical framework of Riemannian geometry~\cite{nojiri2004gravity,allemandi2005dark,manna2023gravity}, providing new insights into the interplay between geometric structures and matter in cosmology and astrophysics. A key distinction emerges between $f(R, L_{m})$ gravity and $f(R)$ theory when matter is incorporated, leading to disparate field equations. Nevertheless, these theories display a remarkable convergence in the vacuum regime, where matter is absent, with their mathematical formulations becoming equivalent~\cite{harko2010f,lobo2015extended}. Further research on $f(R, L_m)$ gravity, documented in~\cite{myrzakulov2024linear,myrzakulova2024investigating,kavya2022constraining,devi2024constraining,garg2025cosmological,garg2024cosmological,doi:10.1142/S0219887823501050}, has significantly advanced our knowledge.
%jaybhaye2022cosmology
\vspace{0.2cm}\\
%Motivated by former research efforts, this research explores several cosmological aspects in the outline of the $f(R, L_m)$ gravity model by using data-driven analytical methods. 
This paper is organized into seven sections, described in detail below: In Section (\ref{sec:2}), we give a brief review of the $f(R, L_m)$ gravity theory and explores the field equations as applied to the homogeneous and isotropic FLRW spacetime metric. In Section (\ref{sec:3}), 
%provides the cosmological parameters derived from a new parametrization of the deceleration parameter, used to obtain exact solutions to the field equations.
an analytical expression for the Hubble parameter is derived by adopting a specific parameterization of the deceleration parameter.
Section (\ref{sec:4}), presents a Bayesian statistical analysis, wherein we utilized Markov Chain Monte Carlo (MCMC) simulations to constrain the median values of the free parameters, for Cosmic Chronometer (CC) dataset and a combined (CC+Pantheon) dataset. In Section (\ref{sec:5}), discover two alternative cosmological models based on $f(R, L_m)$ gravity. Section (\ref{sec:6}), examines the evolution of key cosmological parameters, including energy density $(\rho)$, pressure $(p)$, EoS parameter $(\omega)$ and energy conditions, accompanied by their graphical representations. Moreover, we explore all energy conditions and study the progression of cosmographic parameters. We estimate the universe’s age based on this model. Ultimately, the conclusions are presented in section (\ref{sec:7}).  
%{\bf{compare our model through other dark energy circumstances and contain an analysis of shedding light on the age of the current universe.}} \vspace{0.2cm}\\

\section{Exploring the $f(R, L_{m})$ gravity theory}\label{sec:2}
This section provides an overview of the $f(R, L_{m})$ modified gravity theory and presents the action in $f(R, L_m)$ gravity in the form~\cite{harko2010f}.
%{\bf{the action of this gravity can be extracted as ~\cite{harko2010f}}}
\begin{equation}{\label{1}}
	S=\int f(R, L_{m})\sqrt{-g} d^{4}x,
\end{equation}
where $ L_{m} $ represents the Lagrangian density of matter and $ R $ denotes the Ricci scalar. %curvature
\vspace{.2cm}\\
The relationship between the Ricci scalar $(R)$, Ricci tensor ($R_{\mu \nu}$) and the metric tensor ($g^{\mu \nu}$) is outlined below
\begin{equation}{\label{2}}
	R=g^{\mu \nu}R_{\mu \nu}, 
\end{equation}
where $R_{\mu \nu}$  is incorporated into the resulting expression as follows
\begin{equation}{\label{3}}
	R_{\mu \nu}= \partial_{c} \Gamma^{c}_{\mu \nu}-\partial_{\mu} \Gamma^{c}_{c\nu}+\Gamma^{c}_{\mu \nu}\Gamma^{d}_{dc}-\Gamma^{c}_{\nu d} \Gamma^{d}_{\mu c}.
\end{equation}
Here $ \Gamma^{\alpha}_{\beta \gamma} $ serves as the Levi-Civita connection parameters expressed as
\begin{equation}{\label{4}}
	\Gamma^{\alpha}_{\beta \gamma}= \frac{1}{2}g^{\alpha c}\left(\frac{\partial g_{\gamma c}}{\partial x^{\beta}}+\frac{\partial g_{c \beta}}{\partial x^{\gamma}}-\frac{\partial g_{\beta \gamma }}{\partial x^{c}} \right).
\end{equation}
By applying the variational principle to the action (\ref{1}) with respect to the metric tensor $ g_{\mu\nu} $, we obtain the field equation,
\begin{equation}{\label{5}}
	\frac{\partial F}{\partial R}R_{\mu \nu}+(g_{\mu \nu} \square -\nabla_{\mu}\nabla_{\nu})\frac{\partial F }{\partial R}-\frac{1}{2}\left( F-\frac{\partial F}{\partial L_{m}}L_{m}\right)g_{\mu \nu}=\frac{1}{2}\left(\frac{\partial F}{\partial L_{m}}\right)T_{\mu \nu}, 
\end{equation}
where $\square=g^{\mu \nu}\nabla_{\mu}\nabla_{\nu}$, $ F=f(R, L_{m})$ and $T_{\mu \nu}$ represents the energy-momentum tensor of a perfect fluid, which appears as
\begin{equation}{\label{6}}
	T_{\mu \nu}=\frac{-2}{\sqrt{-g}}\frac{\delta(\sqrt{-g}L_{m})}{\delta g^{\mu \nu}}.
\end{equation}
Utilizing these field equations, we establish a connection between the Ricci scalar $(R)$, the trace of the energy-momentum tensor  $(T)$ and the matter Lagrangian density $(L_{m})$ as

\begin{equation}{\label{7}}
	R\left(\frac{\partial F}{\partial R}\right)  +2\left(\frac{\partial F}{\partial L_{m}}L_{m}-F\right)+ 3\square \frac{\partial F}{\partial R}=\frac{1}{2}\left(\frac{\partial F}{\partial L_{m}}\right)T. 
\end{equation}
Now, we have considered $ \square I=\frac{1}{\sqrt{-g}}\partial_{\mu}(\sqrt{-g}g^{\mu \nu} \partial_{\nu}I)$ for any general function I.
\vspace{.2cm}\\
Substituting the covariant derivative with the energy-momentum tensor allows equation (\ref{5}) to be expressed as:
\begin{equation}{\label{8}}
	\nabla^{\mu}T_{\mu \nu}=2\nabla^{\mu} \log\left(\frac{\partial F}{\partial L_{m}}\right) \frac{\partial L_{m}}{\partial g^{\mu \nu}}.
\end{equation} 
%%%%%%%%%%%%%%%%%%%%%%%%%%%%%%
%%%%%%%%%%%%%%%%%%%%%%%%%%%%%% 
%\section{Field equations}\label{sec:3}
%\section{ Motion equations in $f(R, L_{m})$ \text{ gravity } }\label{sec:3}
%{\bf{The $\Lambda$CDM model aa a standard cosmological model illustrates that the universe has a flat curvature geometry.}}
Most of the cosmological observations suggest that at large scale energy density of the present Universe is very close to the energy density of the flat model of the Universe.
To examine the current cosmological model, we adopt the flat FLRW metric~\cite{partridge2004introduction}, which is formulated as:
\begin{equation}{\label{9}}  
	ds^{2}=-dt^{2}+a^{2}(t) \left( dx^{2}+ dy^{2}+ dz^{2}\right),
\end{equation}
where the cosmic expansion scale factor represented as '$a(t)$' is defined at a specific time '$t$'.
\vspace{.2cm}\\
Based on the line element (\ref{9}), the relevant Christoffel symbols with non-zero values can be identified as:

\begin{equation}{\label{10}}
	\Gamma^{0}_{pq}= -\frac{1}{2}g^{00} \  \frac{\partial g_{pq}}{\partial x^{0}}, \  \  \ \ \Gamma^{r}_{0q}=\Gamma^{r}_{q0}= \frac{1}{2}g^{r\lambda} \  \frac{\partial g_{q \lambda}}{\partial x^{0}},
\end{equation}
The indices $ p, \ q \ and \ r$ takes values $1, \ 2 \ and \ 3$ for space coordinate system.
\vspace{.2cm}\\
Using equation (\ref{3}), the non-zero components of the Ricci tensor are expressed as:
%{\bf{By defining equation number (\ref{3}), we obtain the non-vanishing components of the Ricci tensor, which are as follows.From these equations (\ref{3}), the non-vanishing components of the Ricci tensor are calculated as follows:}} 
\begin{equation}{\label{11}}
	R^{0}_{0}=3\frac{\ddot{a}}{a}, \  \ R^{1}_{1}=R^{2}_{2}=R^{3}_{3}=\frac{\ddot{a}}{a}+2\left(\frac{\dot{a}}{a}\right)^{2}.            
\end{equation}
Thus, based on the line element (\ref{9}), the Ricci scalar is found to be:
\begin{equation}{\label{12}}
	R=6 \ \left[ \left(\frac{\dot{a}}{a}\right)^{2}+ \left( \frac{\ddot{a}}{a}\right) \right] = 6 \ \left(\dot{H}+ 2H^{2} \right).           
\end{equation}
The Hubble parameter $H(t)$ is defined as $ H=\frac{\dot{a}}{a} $.
\vspace{.2cm}\\
In the context of a perfect fluid, the stress-energy momentum tensor is represented as:
%for that filling the universe (\ref{9}), 
\begin{equation}{\label{13}}
	T_{\mu \nu} = (\mathit{p} + \rho) u_{\mu} u_{\nu} + p g_{\mu\nu},
\end{equation}
where the parameters $\mathit{p}$ and $\rho$ characterize the isotropic pressure and energy density of the cosmic fluid. The four-velocity components are expressed as  $u^{\mu} = (1, 0, 0, 0)$  and satisfy the relation $u_{\mu} u^{\mu} = -1 $.
\vspace{.2cm}\\
In  $ f(R, L_m)$ gravity, the evolution of the universe is described by the Friedmann equations, which take the following form:
\begin{equation}{\label{14}}
	\frac{1}{2} (F -F_{L_m} L_m - F_R R) + 3H  \dot{F}_R +3H^2 F_R  = \frac{1}{2} F_{L_m} \rho,
\end{equation}
and 
\begin{equation}{\label{15}}
	3H^2 F_R +\dot{H} F_R- \ddot{F}_R - 3H \dot{F}_R + \frac{1}{2} (F_{L_m} L_m -F) = \frac{1}{2} F_{L_m} p.
\end{equation}
\section{Parametric representations of  $q(z)$ }\label{sec:3}
%Generally, the above system of field equations yields only two independent equations, involving four unknown functions: $\rho$, $p$, $ f(R, L_m)$ and the Hubble parameter $H$. 
The deceleration parameter $q(z)$ plays a central role in characterizing the expansion history of the universe. It directly probes the transition between decelerated and accelerated expansion. In this context, various studies have investigated the deceleration parameter using distinct parametric formulations, while others have adopted non-parametric methods to explore its behavior. These approaches have been extensively examined in the literature to address key challenges in cosmological research, including the initial singularity, the persistent decelerated expansion issue, the horizon problem, the Hubble tension, and related concerns~\cite{banerjee2005acceleration,cunha2008transition}.
%A flexible and well-behaved parameterization of $q(z)$ is therefore critical for model-independent analyses of cosmological data.
We consider the following form:
%~\cite{jesus2018model}:
\begin{equation}{\label{16}}
	q(z)= -1 + \frac{(1+z)(a+2bz)}{(1+az+bz^{2})}.
\end{equation}
The parameters $a$ and $b$ are free model parameters. 
%whose values are constrained by fitting to observational data.
The time derivative of the Hubble parameter satisfies the relation $\dot{H}=-(1+q(z))H^{2}$. Consequently, an integral relation can be derived that links the Hubble parameter to the deceleration parameter:
\begin{equation}{\label{17}}
	H(z)= H_{0} \exp \left[\int_0^z (1 + q(x)) \frac{dx}{1+x} \right],
\end{equation}
where $x$ denotes a variable of integration, $H_{0}$ is the Hubble constant at the present epoch (at $z=0$) and $q(x)$ is the deceleration parameter. 
Substituting equation (\ref{16}) into equation (\ref{17}), we derive the expression for the Hubble parameter as a quadratic function of redshift $z$.
\begin{equation}{\label{18}}
	H(z)= H_{0}(1+az+bz^{2}),
\end{equation}
here $H_{0}$ denotes the present value of the Hubble parameter.
%As a result, we obtain a quadratic expansion in redshift for the normalized Hubble parameter, a functional form commonly adopted in cosmological reconstructions due to its analytical simplicity and wide applicability across observational datasets.
%Our objective is to explore the implications of this scenario for cosmic evolution, using modern observational data as a testing ground. To analyze the behavior of the cosmological observables, the next step involves computing the best-fit estimates for the parameters $a$ and $b$. This will be achieved by combining empirical data from the CC and CC+Pantheon datasets, ensuring a more comprehensive validation of the model.

%%%%%%%%%%%%%%%%%%%%%%%%%%%%%%%%%%%%%%%%%%%%%%%%%%%%%%%%%%%%%%%%%%%%%%%%%%%%%%%%%
\section{Observational constraints on model and results}\label{sec:4}
In this work, we employ a Bayesian statistical framework to evaluate the compatibility of the current cosmological model with observational datasets. For constraining the cosmological parameters $H_{0}$, $a$ and $b$, we consider observational data from the Cosmic Chronometer (CC) sample and the combined CC+Pantheon dataset. We perform model fitting using the  $\chi^{2}$ minimization approach, integrated with the Markov Chain Monte Carlo (MCMC) sampling technique, implemented through the emcee Python library~\cite{foreman2013emcee}.
%\vspace{0.3cm}\\
%In this part of our work, we investigate the Hubble parameters that are compatible with the CC i.e. Cosmic Chronometer and CC+Pantheon i.e. the collective data set of the Cosmic Chronometer and Pantheon nomenclature. 
%\vspace{0.2cm}\\
%{\bf{The Markov Chain Monte Carlo (MCMC) process is executed by using the module of Python library ~\cite{foreman2013emcee} of the Python library~\cite{foreman2013emcee}. By using this method our intent is to obtain median values of model parameters to investigate the cosmological behaviour within the model with $\chi^{2}$ minimization and Bayesian statistical techniques. Several cosmological investigations have been situated in astrophysics and cosmology engaging several parameter sets \cite{mandal2024late, mandal2023cosmic}.
\subsection{The Cosmic chronometer dataset}\label{sec:4.1}
\begin{center}
	\begin{figure}
		\includegraphics[width=15.5cm, height=8.5cm]{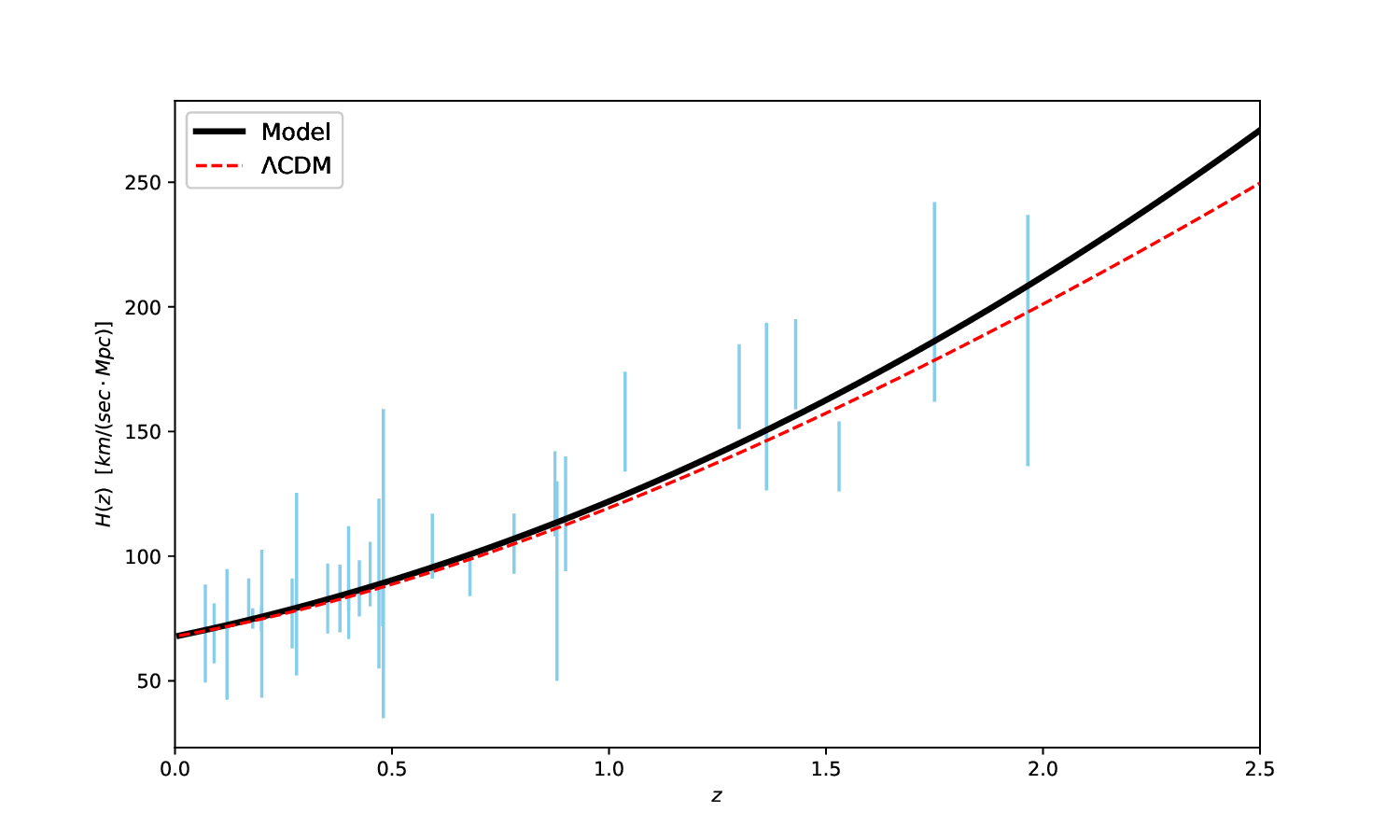}
		\caption{The best-fit $H(z)$ curve with $\mathit{z} $ for the proposed model in comparison to the $\Lambda CDM$ model.} 
		\label{fig:1}
	\end{figure}
\end{center}

	Here, we focus on examining the observational predictions of our cosmological model. We analyze a dataset comprising $31$ cosmic chronometer (CC) measurements~\cite{simon2005constraints,sharov2018predictions}, obtained through the differential age (DA) method applied to galaxies within the redshift range $0.07 \leq z \leq 1.965$~\cite{stern2010cosmic,moresco2015raising}. This analysis is dedicated to calculating the median values of the model parameters. Based on the fundamental principle established by Jimenez and Loeb~\cite{jimenez2002constraining}, the connection between the Hubble parameter $(H(z))$, cosmic-time $(t)$, and redshift $(z)$ is given by the following relation: $H(z)=-\frac{1}{(1+z)}\frac{dz}{dt}$. To estimate the parameters  $H_{0}$, $a$ and $b$, and apply a standard statistical approach involving the minimization of the chi-squared $(\chi^{2})$  function, which is mathematically equivalent to maximizing the likelihood function \cite{mandal2024late, mandal2023cosmic}.
	% This can be statistically signified as:
	\begin{equation}{\label{19}}
	\chi^{2}_{CC}(\theta)=\sum_{i=1}^{i=31} \frac{[H_{th}(\theta,z_{i})-H_{obs}(z_{i})]^{2} }{ \sigma^{2}_{H(z_{i})}}.   
	\end{equation} 
	\vspace{0.1cm}\\
	The symbols $H_{th}$, $H_{obs}$ and  $\sigma_{H}$ denote the theoretical Hubble value, the observed value, and the standard error in the observation, respectively.
	\vspace{0.1cm}\\
	Figure $(\ref{fig:1})$ illustrates the residuals between the cosmic chronometer (CC) observational data and the best-fit curve derived from the Hubble parameter expressions in equation (\ref{18}).
	%\vspace{0.3cm}\\
	%{\bf{Figure $(\ref{fig:1})$ explains the error bars intended for the 31 CC data points regarding the Hubble parameter, as defined in equation (\ref{21}), $H(z)$ empowering the comparison amid $\Lambda$ cold dark matter as the $(\Lambda CDM)$ models and the current model.}} 
	%In addition, Figure $(\ref{fig:2})$ presents a contour plot displaying the $1\sigma$ and $2\sigma$  confidence regions for the median values of $H_{0}$, $a$ and $b$ as constrained by the CC dataset.
	%%%%%%%%%%%%%%%%%%%%%%%%%%%%%%%%%%%%%%%%%%%%%%%%%%%%%%%%%%%
%%%%%%%%%%%%%%%%%%%%%%%%%%%%%%%%%%%%
%%%%%%%%%%%%%%%%%%%%%%%%%%%%%%%%%%%%%%%

%\begin{center}
%	\begin{figure}
	%	\includegraphics[width=18.5cm, height=19.5cm]{FORCCDATA1.eps}
	%	\caption{ $1\sigma$ and $2\sigma$ marginalized contours map with median values of $H_{0}$,  $a$ and $b$ for CC dataset.}
	%	\label{fig:2}
%	\end{figure}
%\end{center}
%%%%%%%%%%%%%%%%%%%%%%%%%%%%%%%%%%%%%%%%%%%%%%%%%%%%%%%%%%%
%%%%%%%%%%%%%%%%%%%%%%%%%%%%%%%%%%%%%%%%%%%%%%%%%%%%%%%%%%%
\subsection{The Pantheon dataset}\label{sec:4.2}
\begin{center}
	\begin{figure}
		\includegraphics[width=18.5cm, height=18.5cm]{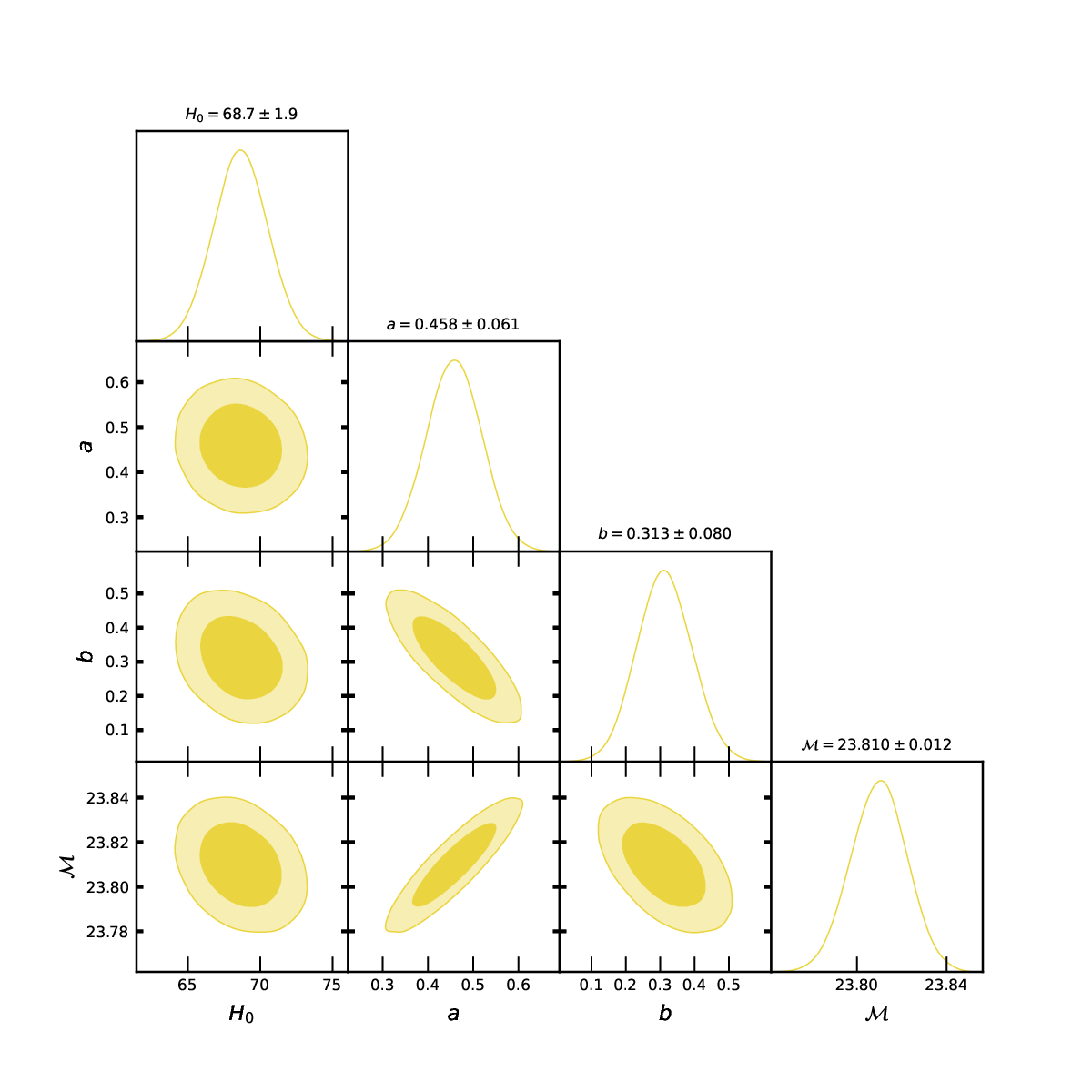}
		\caption{$1D$ and $2D$  marginalized contours map for $H_{0}$, $a$, $b$ and $\mathcal{M}$ using the Joint CC+Pantheon dataset.}
		\label{fig:2}
	\end{figure}
\end{center}

%The most recent release of the Pantheon Type Ia supernova (SNIa) dataset has provided updated observational values for use in cosmological analyses. The dataset comprises $1048$ data points and includes a total of $4010$ individual observations across the redshift range $0.01 < z < 2.26$, as reported in Ref.~\cite{scolnic2018complete}. This compilation includes two subsets from Pan-STARRS1~\cite{scolnic2018complete}: the Medium Deep Survey and the Pan-STARRS1 Light Curve Survey. A significant portion of the data supporting this research originates from the SDSS, SNLS Deep Survey, numerous low-redshift supernova surveys, and HST observations. The study is further supported by findings from the CfA1–CfA4 series~\cite{riess1999bvri,hicken2009improved}, as well as complementary datasets from SDSS~\cite{sako2018data}, SNLS~\cite{guy2010supernova}, and the Carnegie Supernova Project (CSP)~\cite{contreras2010carnegie}, all of which strengthen the Type Ia supernova analysis.
We employ the Pantheon compilation, which consists of 1048 Type Ia supernovae (SNIa) data points covering the redshift range $0.01 < z < 2.26$, as reported in Ref.~\cite{scolnic2018complete}. This comprehensive dataset is assembled from various high-quality surveys, including CfA1–CfA4 series~\cite{riess1999bvri,hicken2009improved}, the Pan-STARRS1 Medium Deep Survey~\cite{scolnic2018complete}, SDSS~\cite{sako2018data}, SNLS~\cite{guy2010supernova}, and the Carnegie Supernova Project (CSP)~\cite{contreras2010carnegie}.
%\vspace{.2cm}\\
%{\bf{The theoretically expected superficial dimension is specified by $\mu_{th}$(z).
In the MCMC analysis utilizing the Pantheon dataset, the theoretically predicted apparent magnitude 
$\mu_{th}(z)$ is expressed as follows:
\begin{equation}{\label{20}}
\mu_{th}(z)=25+5log_{10}\left[\frac{d_{L}(z)}{Mpc}\right]+M,
\end{equation}
where $M$ denotes the absolute magnitude. Additionally, the luminosity distance $d_{L}(z)$, which carries the dimension of length, can be defined as~\cite{odintsov2018cosmological}
\begin{equation}{\label{21}}
	d_{L}(z)=c(1+z)\int_{0}^{z}\frac{dz'}{H(z')},
\end{equation}
where the parameter $z$ corresponds to the redshift of Type Ia supernovae (SNIa) as observed in the cosmic microwave background (CMB) rest frame, while  $c$ is the speed of light. The luminosity distance $(d_L(z))$ is often replaced by the dimensionless Hubble-free luminosity distance, which can be written as $D_{L}(z) \equiv H_{0}d_{L}(z)/c$.
%{\bf{The Hubble free luminosity distance$(D_{L}(z) \equiv H_{0}d_{L}(z)/c)$ use instead of luminosity distance $(d_{L})$.}} 
Alternatively, equation (\ref{20}) may be rewritten as:
\begin{equation}{\label{22}}
	\mu_{th}(z)=25+5log_{10}\left[D_{L}(z)\right]+5log_{10}\left[\frac{c/H_{0}}{Mpc}\right]+M. 
\end{equation}
The parameters $M$ and $H_{0}$ can also be used to define a new parameter such as $\mathcal{M}$, which can be expressed as
\begin{equation}{\label{23}}
	\mathcal{M}\equiv M+25+5log_{10} \left[\frac{c/H_{0}}{Mpc}\right]=M-5log_{10}(h)+42.38, 
\end{equation}
where $H_{0}=h \times 100$ $ [Km/(s. Mpc)]$. For the MCMC analysis, we utilize these parameters in conjunction with the corresponding $\chi^{2}$ function for the Pantheon dataset as~\cite{asvesta2022observational}
\begin{equation}{\label{24}}
	\chi^{2}_{P}= \nabla \mu_{i}C^{-1}_{ij}\nabla \mu_{j},
\end{equation}
where $\nabla \mu_{i}=\mu_{obs}(z_{i})-\mu_{th}(z_{i})$, and $C_{ij}^{-1}$ denotes the inverse of the covariance matrix, while $\mu_{th}$  is defined by equation (\ref{22}). 
\vspace{.2cm}\\
The luminosity distance depends on the behavior of the Hubble parameter. Earlier, we employed the emcee package~\cite{foreman2013emcee} along with equation (\ref{18}) to perform maximum likelihood estimation (MLE) using the combined CC+Pantheon dataset. The joint  $\chi^{2}$ statistic used for MLE is constructed as $\chi^{2}_{CC}+\chi^{2}_{P}$. In Fig. $(\ref{fig:2})$, we present the $1D$ and $2D$ likelihood contours and corresponding 1D posterior distributions derived from the MCMC sampling of the CC+Pantheon joint dataset.
%\vspace{.2cm}\\
%For the MCMC study, we utilize 40000 iterations (steps) and $32$ random chains(walkers) for the CC data set. For the CC data set, We pick uniform priors on $h_{0}$\,$w_{0}$ \ and $n$. The ranges we use are $40 < h_{0} < 90$,\ $ -3 < w_{0} < 0$ \ and\  $0.8 < n < 1.2$. We utilized 10000 iterations(steps) and $48$ random chains(walkers) for the MCMC observations of the joint dataset. For the joint data set, we take uniform priors on $h_{0}$\,$w_{0}$ \ and $n$ and $\mathcal{M}$. The ranges we use are $40 < h_{0}< 90,\ -3 < w_{0} < 0 ,\ 0.8 < n < 1.2 $,and \  $ 23.5<\mathcal{M}<23.95$.
%\vspace{.2cm}\\
The Table(\ref{table:1}) provides a concise overview of the median values for the model parameters.

\begin{table}[htbp]
\centering
\tiny{
\begin{tabular}{|c|c|c|c|c|c|c|c|c|c|}
\hline
Dataset & $H_{0}$[Km/(s.Mpc)] & $a$ & $b$ & $\mathcal{M}$ & $q_{0}$ & $z_{t}$ & $j_{0}$ & $s_{0}$ & $t_{0}$[Gyr] \\
\hline
CC & $67.8^{+1.7}_{-1.7}$ & $0.54^{+0.11}_{-0.11}$ & $0.265^{+0.058}_{-0.060}$ & - & $-0.4599$ & $0.655$ & $0.7416$ & $-0.5175$ & $12.97^{+1.63}_{-1.25}$ \\
\hline
CC+Pantheon  & $68.7^{+1.9}_{-1.9}$  & $0.458^{+0.061}_{-0.061}$ &  $0.313^{+0.080}_{-0.080}$ &  $23.810^{+0.012}_{-0.012}$ & $-0.542$ & $0.655$ & $0.9197$ & $-0.3616$ & $12.81^{+1.51}_{-1.17}$ \\
\hline
\end{tabular} }
\caption{Best-fit values of the model parameters using MCMC, along with $q_{0}$, $z_{t}$, $j_{0}$, $s_{0}$ and $t_{0}$.}
\label{table:1}
\end{table}
%%%%%%%%%%%%%%%%%%%%%%%%%%%%%%%%%%%%%%%%%%%%%%%%%%%%%%%%%%%
%%%%%%%%%%%%%%%%%%%%%%%%%%%%%%%%%%%%%%%%%%%%%%%%%%%%%%%%%%%
%\section{Cosmological models based on $f(R, L_{m})$ gravity}\label{sec:6}
\section{Cosmological analysis of $f(R, L_{m})$ gravity models}\label{sec:5}
This theoretical investigation explores two nonlinear $f(R, L_{m})$ models with the free parameter $\eta$, expressed in the forms $f(R, L_{m})$ = $\frac{R}{2}+(1+\eta R) L_{m}$ and $f(R, L_{m})$ = $\frac{R}{2}+ L_{m}^{\eta}$, to address the modifications to gravitational dynamics. At the core of these models is a comprehensive $f(R, L_{m})$ theory, featuring arbitrary curvature-matter coupling, formally written as        $f(R, L_{m})= f_{1}(R)+f_{2}(R)G(L_{m})$~\cite{harko2014generalized}. 
%Cosmologists are showing growing interest in both minimal and nonminimal coupling frameworks, particularly in the context of diverse modified gravity models.
An increasing focus within cosmology is directed at exploring minimal and nonminimal coupling mechanisms, especially as they arise in various formulations of modified gravity.
Notably, the general $f(R, L_{m})$ framework described above is broad enough to include cases of both minimal and non-minimal curvature-matter coupling. Our analysis is intended to provide a clearer understanding of how minimal and non-minimal coupling mechanisms affect the evolution of dark energy in cosmological theories. 

\subsection{Model (I) : $f(R, L_{m}) = \frac{R}{2}+(1+\eta R) L_{m}$ }\label{sec:5.1}
To assess the cosmological consequences of modified gravity, we begin by analyzing a non-minimal coupling scenario guided by the functional form of $f(R, L_{m})$ in~\cite{lobato2021neutron,myrzakulova2024investigating}. The selection of this non-minimal coupling scenario is inspired by numerous prior studies that have investigated analogous formulations,
\begin{equation}{\label{25}}
	f(R, L_{m}) = \frac{R}{2}+(1+\eta R) L_{m}.
\end{equation}
The symbol $\eta$ corresponds to a free parameter whose value is constrained observationally. Examining the role of the coupling parameter within distinct cosmological frameworks remains a key aspect of ongoing theoretical investigations. It is worth noting that setting $\eta=0$ recovers the standard Friedmann equations from general relativity. Here, the model takes $L_{m}=\rho$ in the $f(R, L_{m})$ formulation~\cite{harko2015gravitational}. The Friedmann equations given in (\ref{14}) and (\ref{15}) can be rewritten in the following form:
\begin{equation}{\label{26}}
	3H^{2}(6 \eta \rho-1)+ 12\eta \dot{H}\rho+\rho=0,
\end{equation}
\begin{equation}{\label{27}}
	3H^{2}(-2\eta \rho+4\eta p+1)+ \dot{H}(-2\eta \rho+6\eta p+2)+p =0.
\end{equation}

\subsection{Model (II) : $f(R, L_{m}) = \frac{R}{2}+ L_{m}^{\eta}$ }\label{sec:5.2}
In order to investigate the evolution of dark energy (DE), we consider a particular form of the minimal $f(R, L_{m})$ function as presented in~\cite{harko2014generalized}. The present investigation of minimal coupling is inspired by the pioneering study of Bose et al.~\cite{bose2022analytic}, situated in the theoretical setting of $f(R, T)$ gravity,
\begin{equation}{\label{28}}
f(R, L_{m}) = \frac{R}{2}+ L_{m}^{\eta},
\end{equation}
where $\eta$ corresponds to a free model parameter. It is worth mentioning that setting $\eta=0$ yields the conventional Friedmann equations of General Relativity. Assuming $L_{m}=\rho$ in the chosen $f(R, L_{m})$ model~\cite{harko2015gravitational}, equations (\ref{14}) and (\ref{15}), corresponding to the Friedmann system, take the following form:
\begin{equation}{\label{29}}
	3H^{2}= (2 \eta -1) \rho^{\eta},
\end{equation}
\begin{equation}{\label{30}}
2 \dot{H}+3H^{2}=[(\eta -1)\rho - \eta p] \rho^{\eta-1}.
\end{equation}

%%%%%%%%%%%%%%%%%%%%%%%%%%%%%%%%%%%%%%%%%%%%%%%%%%%%%%%%%%%

\section{Analysis of the model's physical and dynamical characteristics}\label{sec:6}

\subsection{ Deceleration parameter }\label{sec:6.1}
\begin{center}
	\begin{figure}
		\includegraphics[width=15.5cm, height=8.5cm]{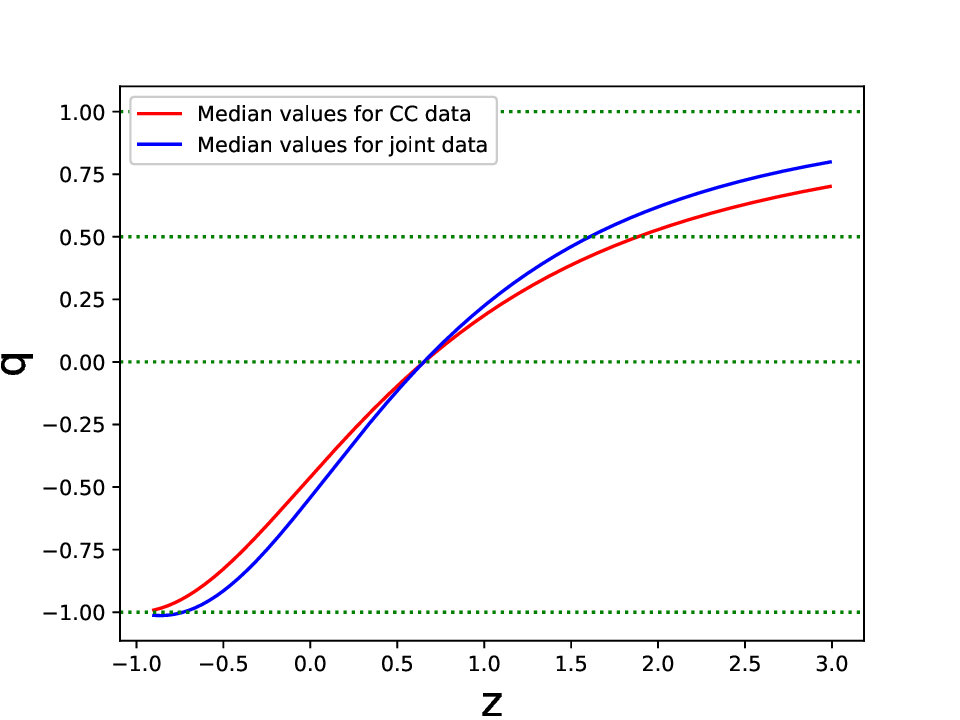}
		\caption{The deceleration parameter$(q)$ versus $z$.} 
		\label{fig:3}
	\end{figure}
\end{center}

%\vspace{0.2cm}\\
%To be specific, the deceleration parameter $q=\frac{-\dot{a}}{aH^{2}} $ may be used to characterize the pace of universe expansion. As an alternative, the relationship may be stated as 
One of the fundamental quantities governing cosmic expansion is the deceleration parameter $(q)$. It shows how the universe behaves. Different values of the deceleration parameter are employed to characterize the universe’s expansion, where $ q<0 $ corresponds to an accelerated expansion phase, and $ q>0 $ indicates a decelerated phase. If the deceleration parameter falls below $-1$, it signals a transition into a super-accelerated expansion phase of the universe. Observed expansion behaviors of the universe align with characteristic $q$ values namely $-1$, $\frac{1}{2}$ and $1$ which correspond to de Sitter, matter-dominated, and radiation-dominated regimes, respectively. Equation (\ref{16}) provides the explicit expression for the deceleration parameter $q(z)$ used in this analysis.\\
Figure $(\ref{fig:3})$ illustrates the deceleration parameter analysis for both the CC and joint data sets. Figure $(\ref{fig:3})$ captures the cosmic transition from an early decelerated expansion era to the current phase of accelerated expansion.
The redshift at which the transition occurs is $z=0.655$ for both the CC and joint data sets, based on the median values of the model parameters. The present-day deceleration parameter is found to be $q_{0}=-0.4599$ for the CC data set and $q_{0}=-0.542$ for the joint data set. The negative values indicate that the universe is presently experiencing accelerated expansion at redshift $z=0$.
The trend of $q$ in Fig.$(\ref{fig:3})$ suggests that the model captures a matter-dominated phase in the early universe, where $q$ tends toward $\frac{1}{2}$. It then describes the current accelerated expansion $(q<0) $ and eventually the model predicts convergence to a de Sitter phase with $q=-1$ as $z \to -1$.
This behavior is further supported by the deceleration parameter, which confirms the current acceleration phase.

%These negative values prove that the universe's expansion is presently accelerating.
%\begin{figure}[!htb]
%\captionsetup{skip=0.4\baselineskip,size=footnotesize}
%   \begin{minipage}{0.40\textwidth}
%     \centering
%     \includegraphics[width=9.0 cm,height=7.5cm]{DECELERATION.eps}
%\caption{The deceleration parameter$(q)$ versus $z$.}
%\label{fig:4}
%    \end{minipage}\hfill
%   \begin{minipage}{0.40\textwidth}
%     \centering
%     \includegraphics[width=9.0 cm,height=7.5cm]{EOS.eps}
%  \caption{Equation of State parameter (given in Eq. (\ref{26})) versus $z$.}
%\label{fig:5}
%   \end{minipage}
%\end{figure}

\subsection{Observational Study of Energy Density and Pressure}\label{sec:6.2}
\begin{figure}[!htb]
\captionsetup{skip=0.4\baselineskip,size=footnotesize}
   \begin{minipage}{0.50\textwidth}
     \centering
     \includegraphics[width=9.0 cm,height=7.5cm]{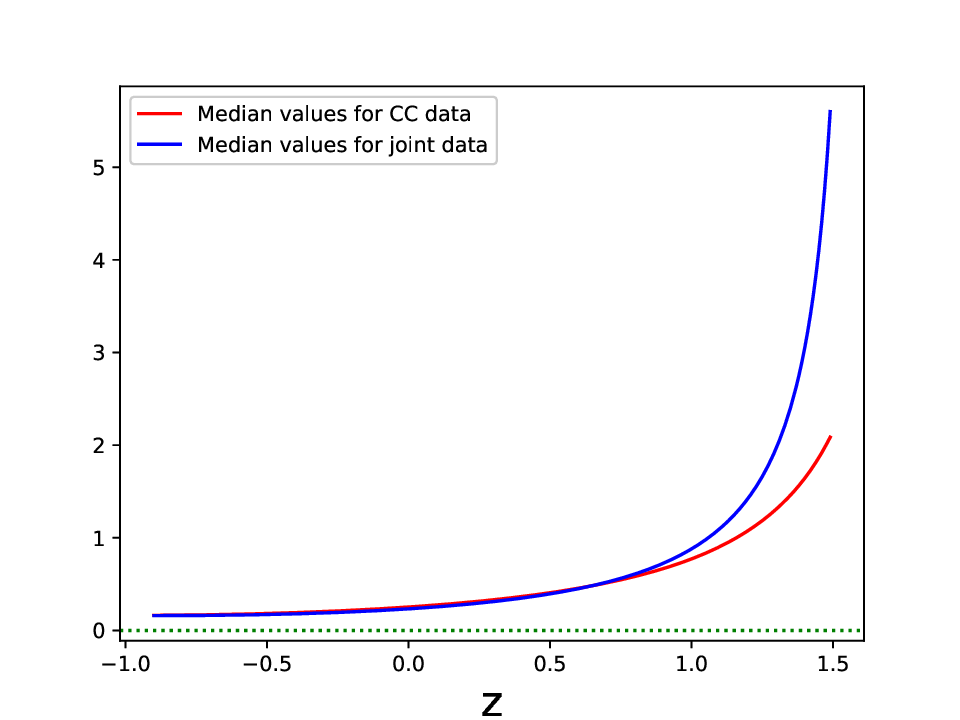}
\caption{Energy density$(\rho)$ versus $z$ for model(I).}
\label{fig:4}
    \end{minipage}\hfill
   \begin{minipage}{0.50\textwidth}
     \centering
     \includegraphics[width=9.0 cm,height=7.5cm]{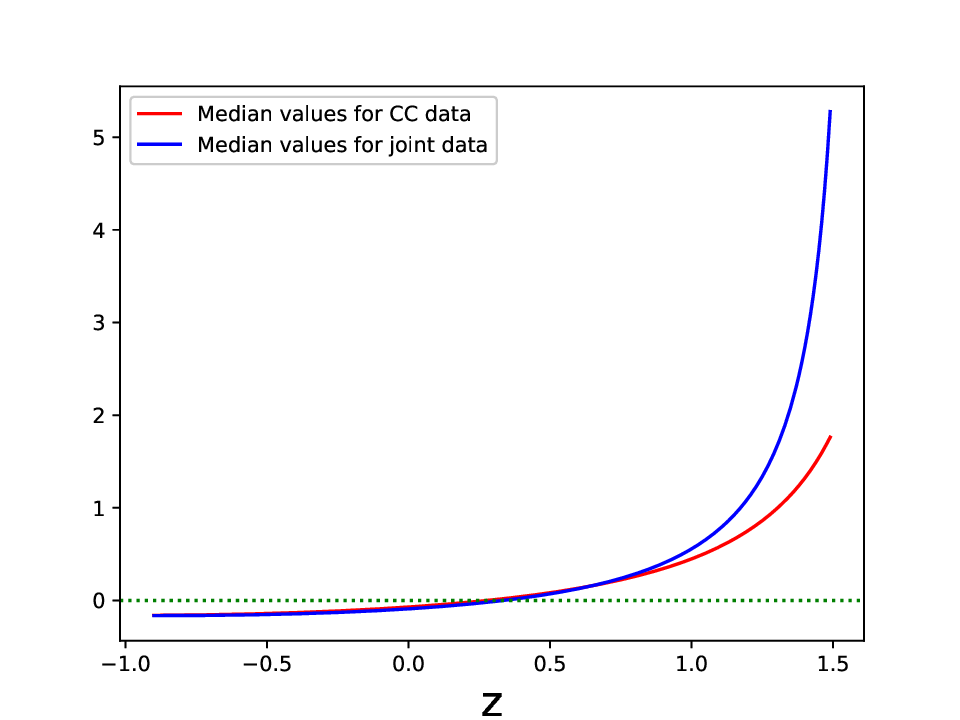}
  \caption{Pressure($p$) versus $z$ for model(I).}
\label{fig:5}
   \end{minipage}
\end{figure}

\begin{figure}[!htb]
	\captionsetup{skip=0.4\baselineskip,size=footnotesize}
	\begin{minipage}{0.50\textwidth}
		\centering
		\includegraphics[width=9.0 cm,height=7.5cm]{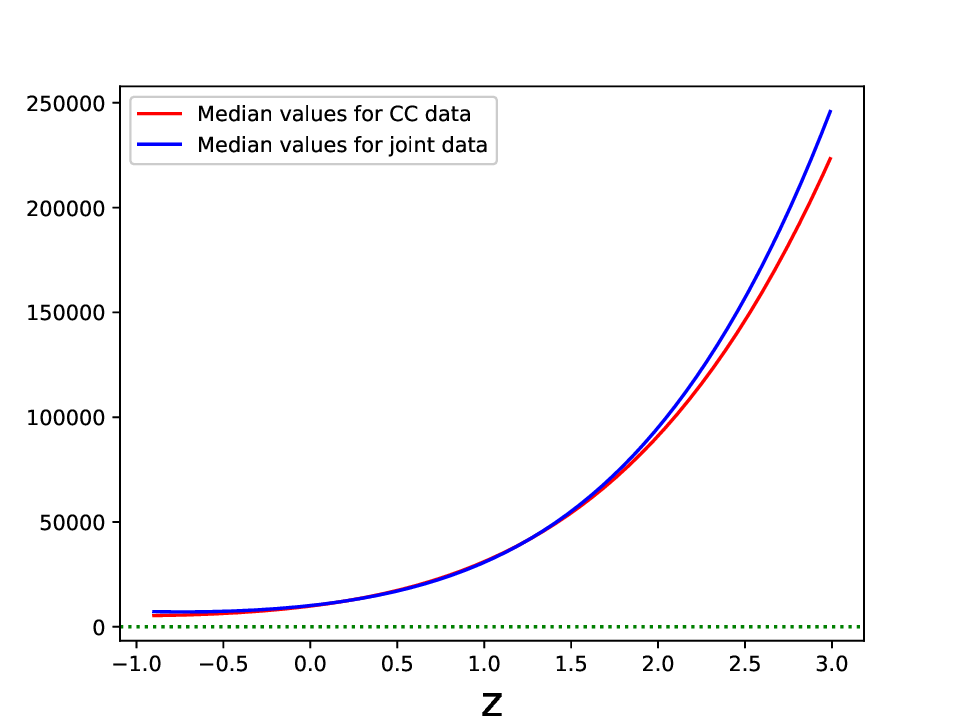}
		\caption{Energy density$(\rho)$ versus $z$ for model(II).}
		\label{fig:6}
	\end{minipage}\hfill
	\begin{minipage}{0.50\textwidth}
		\centering
		\includegraphics[width=9.0 cm,height=7.5cm]{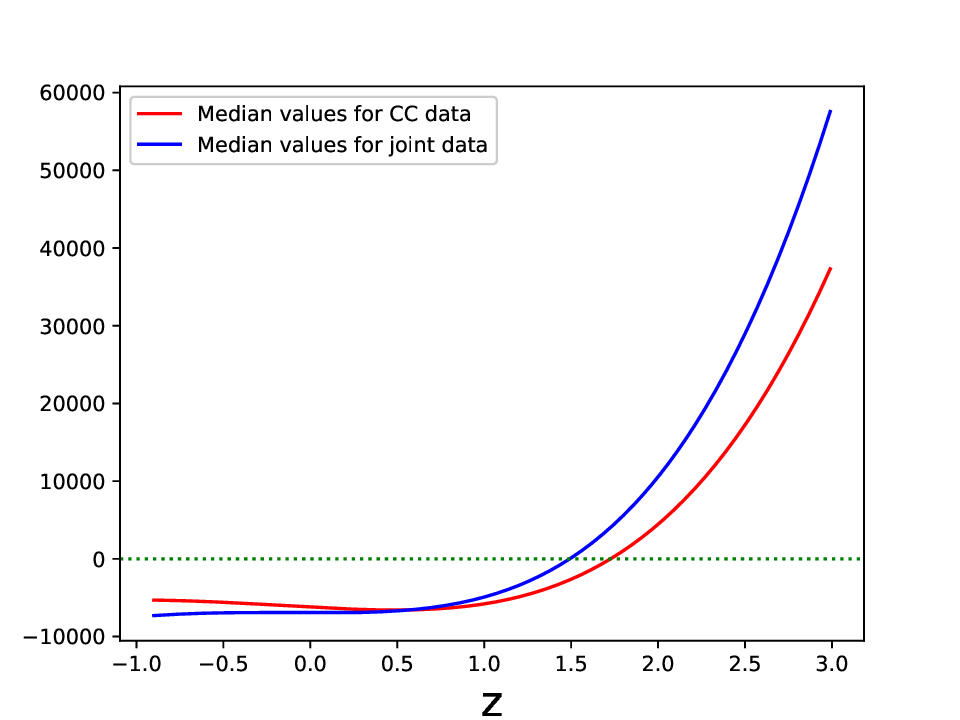}
		\caption{Pressure($p$) versus $z$ for model(II).}
		\label{fig:7}
	\end{minipage}
\end{figure}

Throughout the cosmic expansion, the energy density remains positive, whereas the pressure is likely to have turned negative in the recent epoch. The negative pressure observed since the recent past may be responsible for driving the accelerated expansion of the universe in this model. During the earlier decelerating phase, the expansion history is characterized by a positive but decreasing energy density. As the universe transitions from deceleration to acceleration, the energy density remains positive, while the pressure turns negative, an effect attributed to the increasing dominance of dark energy. 
%This behavior is particularly evident during the transition from decelerated to accelerated expansion, where dark energy becomes dominant, resulting in negative pressure while maintaining positive energy density.
%\vspace{0.2cm}\\
%{\bf{Although pressure goes through a transition from positive to negative values is the scenario of an accelerated universe. Whereas positive and decreasing energy density is the scenario of the decelerating age. As a result of the governance of dark energy, the energy density has remained positive and pressure turned into negative while shifting deceleration to the accelerational phase.}}
\vspace{.2cm}\\
According to equations (\ref{26}), (\ref{27}), (\ref{29}) and (\ref{30}) the expressions for energy density ($\rho$) and pressure (p) are derived using the Hubble parameter and its derivative with respect to time.
\begin{equation}{\label{31}}
\rho(z) =  \frac{3H_{0}^{2} (1+az+bz^{2})^{2}}{H_{0}^{2}(1+az+bz^{2})\left[18\eta(1+az+bz^{2})- 12 \eta(1+z)(a+2bz)\right] +1} ~~~~~~~~~~~~~~~~~~~~~~~~ \qquad \qquad Model(I)
\end{equation}
%%%%%%%%%%%%%%%%%%%%%%%%%%%%%%%%%%%%%%%%%%%%
%{\small\begin{align}
% \mathit{p}(z) =&\frac{3H_{0}^{2}(1+az+bz^{2})^{2} - 2H_{0}^{2}(1+z)(a+2bz)(1+az+bz^{2})}{ \left[6\eta H_{0}^{2}(2(1+az+bz^{2})^{2} - (1+z)(a+2bz)(1+az+bz^{2})) + 1 \right]}-\nonumber\\
%& \frac{H_{0}(1+az+bz^{2})[6\eta H_{0}^{3}(2(1+az+bz^{2})-(1+z)(a+2bz)) (3 (1+az+bz^{2})^{2} - 4(1+z) (a+2bz) (1+az+bz^{2}))] }{\left[18\eta H_{0}^{2}(1+az+bz^{2})^{2}-12\eta H_{0}^{2} (1+z)(a+2bz)(1+az+bz^{2}) + 1 \right] \left[6\eta H_{0}^{2}(2(1+az+bz^{2})^{2} - (1+z)(a+2bz)(1+az+bz^{2})) + 1 \right] }
%\end{align}}
%%%%%%%%%%%%%%%%%%%%%%%%%%%%%%%%%%%%%%%%%%%
%\begin{align}
%	\rho(z) = &  \frac{3H_{0}^{2} (1+az+bz^{2})^{2}}{H_{0}^{2}(1+az+bz^{2})\left[18\eta(1+az+bz^{2})- 12 \eta(1+z)(a+2bz)\right] +1} & Model(I)
%\end{align}	
%\begin{align}
%\mathit{p}(z) =  & -\frac{H_{0}E[6\eta H_{0}^{3}(2E-(1+z)(a+2bz)) (3 E^{2} - 4(1+z) (a+2bz) E)] }{\left[18\eta H_{0}^{2}E^{2}-12\eta H_{0}^{2} (1+z)(a+2bz)E + 1 \right] \left[6\eta H_{0}^{2}(2E^{2} - (1+z)(a+2bz)E) + 1 \right] } & \nonumber\\
%&+\frac{3H_{0}^{2}E^{2} - 2H_{0}^{2}(1+z)(a+2bz)E}{ \left[18\eta H_{0}^{2}E^{2}-12\eta H_{0}^{2} (1+z)(a+2bz)E + 1 \right] \left[6\eta H_{0}^{2}(2E^{2} - (1+z)(a+2bz)E) + 1 \right]} &  Model(I)
%\end{align}
%where $E(z)= 1+az+bz^{2}$
%%%%%%%%%%%%%%%%%%%%%%%%%%%%%%%%%%
\begin{equation*}
 \mathit{p}(z) =  -\frac{H_{0}E[6\eta H_{0}^{3}(2E-(1+z)(a+2bz)) (3 E^{2} - 4(1+z) (a+2bz) E)] }{\left[18\eta H_{0}^{2}E^{2}-12\eta H_{0}^{2} (1+z)(a+2bz)E + 1 \right] \left[6\eta H_{0}^{2}(2E^{2} - (1+z)(a+2bz)E) + 1 \right] }   ~~~~~~~~~~~~~~~~~~~~~~~~~~~~
\end{equation*} 
\begin{equation}{\label{32}}
\qquad	+\frac{3H_{0}^{2}E^{2} - 2H_{0}^{2}(1+z)(a+2bz)E}{ \left[18\eta H_{0}^{2}E^{2}-12\eta H_{0}^{2} (1+z)(a+2bz)E + 1 \right] \left[6\eta H_{0}^{2}(2E^{2} - (1+z)(a+2bz)E) + 1 \right]} ~~~~~  Model(I)
\end{equation} 
where $E= 1+az+bz^{2}$.
%%%%%%%%%%%%%%%%%%%%%%%%%%%%%%%%%%%%%%%%%%%%%%%%%
%{ \small
%	\begin{equation*}
%		\mathit{p}(z) =  -\frac{H_{0}(1+az+bz^{2})[6\eta H_{0}^{3}(2(1+az+bz^{2})-(1+z)(a+2bz)) (3 (1+az+bz^{2})^{2} - 4(1+z) (a+2bz) (1+az+bz^{2}))] }{\left[18\eta H_{0}^{2}(1+az+bz^{2})^{2}-12\eta H_{0}^{2} (1+z)(a+2bz)(1+az+bz^{2}) + 1 \right] \left[6\eta H_{0}^{2}(2(1+az+bz^{2})^{2} - (1+z)(a+2bz)(1+az+bz^{2})) + 1 \right] }   ~~~~~~~~
%\end{equation*} }
%{ \small
%	\begin{equation}{\label{33}}
%		+\frac{3H_{0}^{2}(1+az+bz^{2})^{2} - 2H_{0}^{2}(1+z)(a+2bz)(1+az+bz^{2})}{ \left[18\eta H_{0}^{2}(1+az+bz^{2})^{2}-12\eta H_{0}^{2} (1+z)(a+2bz)(1+az+bz^{2}) + 1 \right] \left[6\eta H_{0}^{2}(2(1+az+bz^{2})^{2} - (1+z)(a+2bz)(1+az+bz^{2})) + 1 \right]}  ~~~~~~~~~~~~~~~~~~~~~~~~ Model(I)
%\end{equation} }
%%%%%%%%%%%%%%%%%%%%%%%%%%%%%%%%%%%%%%%%%%%%%%%%%
\begin{equation}{\label{33}}
	\rho(z) = \left(\frac{3H_{0}^{2} (1+az+bz^{2})^{2} }{2\eta-1} \right)^{1/\eta} ~~~~~~~~~~~~~~~~~~~~~~~~~~~~~~~~~~~~~~~~~~~~~~~~~~~~~~~~~~~~~~~~~~~~~~ \qquad \qquad \qquad  Model(II)
\end{equation}
\begin{equation}{\label{34}}
\mathit{p}(z) = \frac{\left(\frac{3H_{0}^{2} (1+az+bz^{2})^{2} }{2\eta-1} \right)^{1/\eta} \left(3\eta(1+az+bz^{2}) - (4\eta-2) (1+z)(a+2bz)\right)}{3\eta(1+az+bz^{2})}  ~~~~~ \qquad \qquad \qquad Model(II)
\end{equation} 
In the case of model(I), figures (\ref{fig:4}) and (\ref{fig:5}) illustrate the temporal evolution of the energy density and pressure respectively. Correspondingly, for model(II), figures (\ref{fig:6}) and (\ref{fig:7}) present the evolution of these quantities.
Moreover, the pressure ($\mathit{p}(z)$) component is observed to remain negative during the present epoch and in future evolution, whereas the energy density ($\rho(z)$) maintains a positive value, consistent with the dynamics of an accelerating universe. The negative pressure observed in the late-time Universe may be a driving factor behind its accelerated expansion. This behavior is in agreement with current observational evidence supporting a rapidly expanding cosmos. These findings are in alignment with the characteristic expansion of an accelerating Universe. For the graphical representation in both Model(I) and Model(II), we adopt the value $\eta = 1.03$. \\
Figures (\ref{fig:4}) to (\ref{fig:7}) show that for both models, the energy density increases with redshift$(z)$ meaning it decreases with cosmic time(t) and remains positive throughout the universe's evolution. 
%The consistent of positive energy density at all epochs effectively rules out the occurrence of finite-time future singularities in these models.
At early times (high z), the pressure is positive, reflecting a decelerating expansion phase. As the universe evolves, the pressure gradually becomes negative, particularly in the late-time and present epochs, signifying a transition to an accelerated expansion era. This negative pressure behavior is consistent with dark energy dominance and supports the viability of both models in describing the observed cosmic acceleration.

%{\bf{Those conclusions support the concept of accelerating the expansion of the Universe.}}
%%%%%%%%%%%%%%%%%%%%%%%%%%%%%%%%%%%%%%%%%%%%%%%%%%%
%%%%%%%%%%%%%%%%%%%%%%%%%%%%%%%%%%%%%%%%%%%%%%%%%%%%%%%%%%
\begin{figure}[!htb]
	\captionsetup{skip=0.4\baselineskip,size=footnotesize}
	\begin{minipage}{0.50\textwidth}
		\centering
		\includegraphics[width=9.0 cm,height=7.5cm]{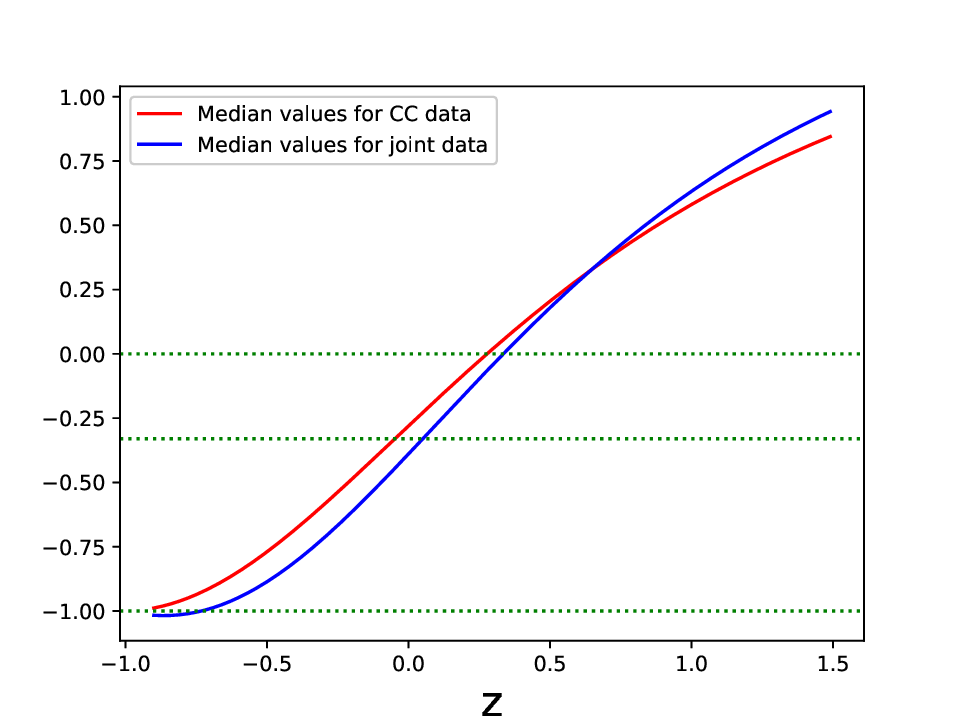}
		\caption{EoS parameter ($\omega$) versus $z$ for model(I).}
		\label{fig:8}
	\end{minipage}\hfill
	\begin{minipage}{0.50\textwidth}
		\centering
		\includegraphics[width=9.0 cm,height=7.5cm]{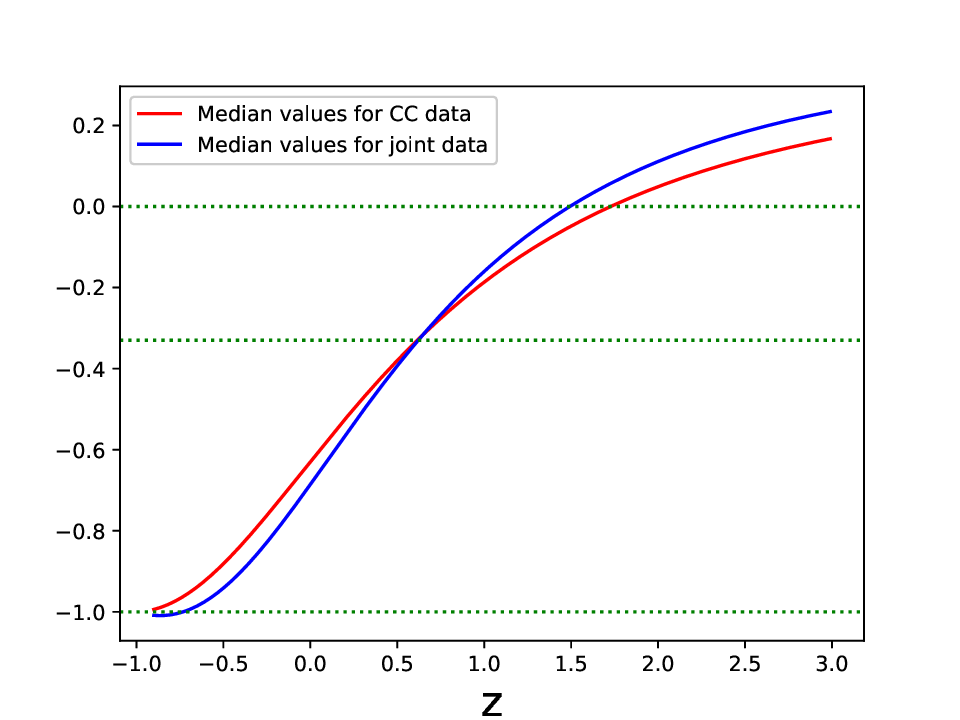}
		\caption{EoS parameter ($\omega$) versus $z$ for model(II).}
		\label{fig:9}
	\end{minipage}
\end{figure}

%%%%%%%%%%%%%%%%%%%%%%%%%%%%%%%%%%%%%%%%%%%%%%%%%%%%%%%%%%%
\subsection{Equation of State (EoS) Parameter}\label{sec:6.3}
The equation of state parameter describes the relationship between pressure $(\mathit{p})$ and energy density $(\rho)$. It plays a key role in characterizing different cosmic phases for instance, $\omega=0$ corresponds to pressureless matter (dust), $\omega=\frac{1}{3}$ represents the radiation-dominated era, and  $\omega=-1$ signifies vacuum energy, associated with the de-Sitter phase in the $\Lambda$CDM model. Moreover, the accelerated expansion era of the Universe is associated with values of the EoS parameter less than $-\frac{1}{3}$. This interval includes the phantom region, where $\omega < -1 $, and the quintessence region, defined by $-1 < \omega < -\frac{1}{3}$. By employing Eqs. (\ref{31}) through (\ref{34}), the EoS parameter $(\omega=\frac{\mathit{p}}{\rho}) $ can be derived in the following manner:
\begin{equation}{\label{35}}
\omega(z)= -1 - \frac{2(1+z)(a+2bz)}{3 (1+az+bz^{2})} \left(\frac{1}{6\eta H_{0}^{2} (1+az+bz^{2}) \left[2(1+az+bz^{2}) - (1+z)(a+2bz)\right] +1 } - 2 \right) ~~Model(I)
\end{equation}
\begin{equation}{\label{36}}
\omega(z)= -1 - \frac{(2-4\eta)(1+z)(a+2bz)}{3\eta (1+az+bz^{2})} ~~~~~~~~~~~~~~~~~~~~~~~~~~~~~~~~~~~~~~~~~~~~~~~~~~~~~~~~~~~~~~~~~~~~~~~~~~~~~~~~~~~~~~~~~~~Model(II)
\end{equation}
Based on the median values of the model parameters, the corresponding evolution of the dark energy EoS parameter is presented in Figures (\ref{fig:8}) and (\ref{fig:9}). For Model (I), the present-day EoS parameter is determined to be $\omega_{0}=-0.2800$ based on the CC dataset, while the joint analysis incorporating both CC and Pantheon data yields $\omega_{0}=-0.3893$. In contrast, Model (II) predicts $\omega_{0} = -0.6295$ using the CC dataset and $\omega_{0} =-0.6857$ from the combined dataset at redshift $z=0$. These results suggest that Model(II) exhibits a stronger late-time acceleration and is more consistent with dark energy-dominated dynamics compared to Model(I). These outcomes indicate that, at present, Model(II) exhibit a quintessence-like dark energy behavior.
Figures (\ref{fig:8}) and (\ref{fig:9}) confirm that these models undergo a matter-dominated expansion in the early universe, characterized by $\omega=0$.
%In the current epoch, the dynamics are driven by a quintessence-like dark energy component with $-1 < \omega < -\frac{1}{3}$ and
In the far future, these models tend toward the $\Lambda$CDM scenario with $\omega=-1$ as $z \to -1$.

%At late times, as $z \to -1$, the models converge toward a cosmological constant-dominated regime, consistent with a de Sitter expansion.
%In Model I, the equation of state $\omega$ goes below $−1$ in the future, which means the model shows quintom behavior—where dark energy becomes stronger over time, leading to faster expansion.
%In Model II, there may be an exchange of energy between dark energy and dark matter. This could slow down the Universe’s expansion in the future, making it an interesting feature of the model.

\subsection{Frameworks for Energy Conditions}\label{sec:6.4}
\begin{figure}[!htb]
\captionsetup{skip=0.4\baselineskip,size=footnotesize}
   \begin{minipage}{0.50\textwidth}
     \centering
\includegraphics[width=8.2cm,height=7cm]{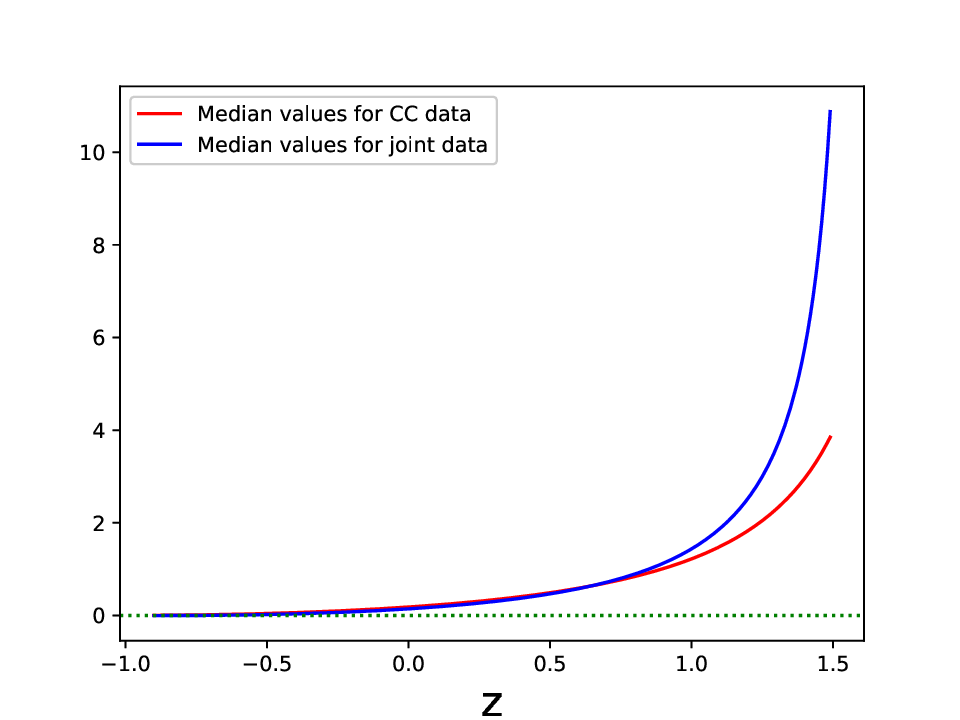}
\caption{For Model(I): $(\rho + p)$ versus $\mathit{z}$.}
\label{fig:10}
    \end{minipage}\hfill
   \begin{minipage}{0.50\textwidth}
     \centering
     \includegraphics[width=8.2cm,height=7cm]{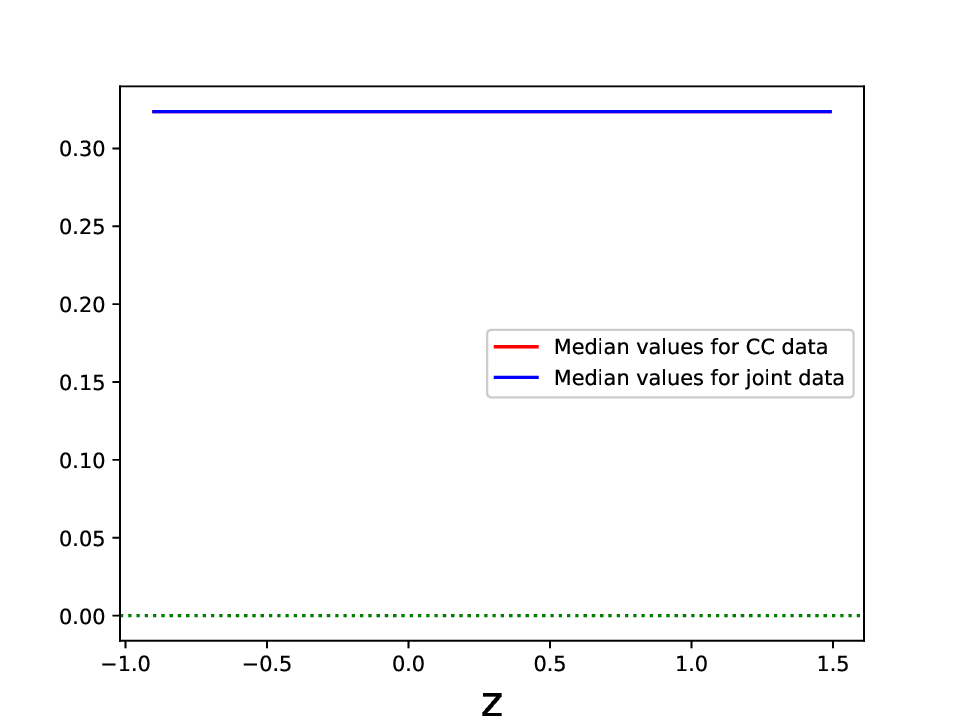}
  \caption{For Model(I): $(\rho - p)$ versus $\mathit{z}$.}
\label{fig:11}
   \end{minipage}
\end{figure}

\begin{figure}[!htb]
	\captionsetup{skip=0.4\baselineskip,size=footnotesize}
	\begin{minipage}{0.50\textwidth}
		\centering
		\includegraphics[width=8.2cm,height=7cm]{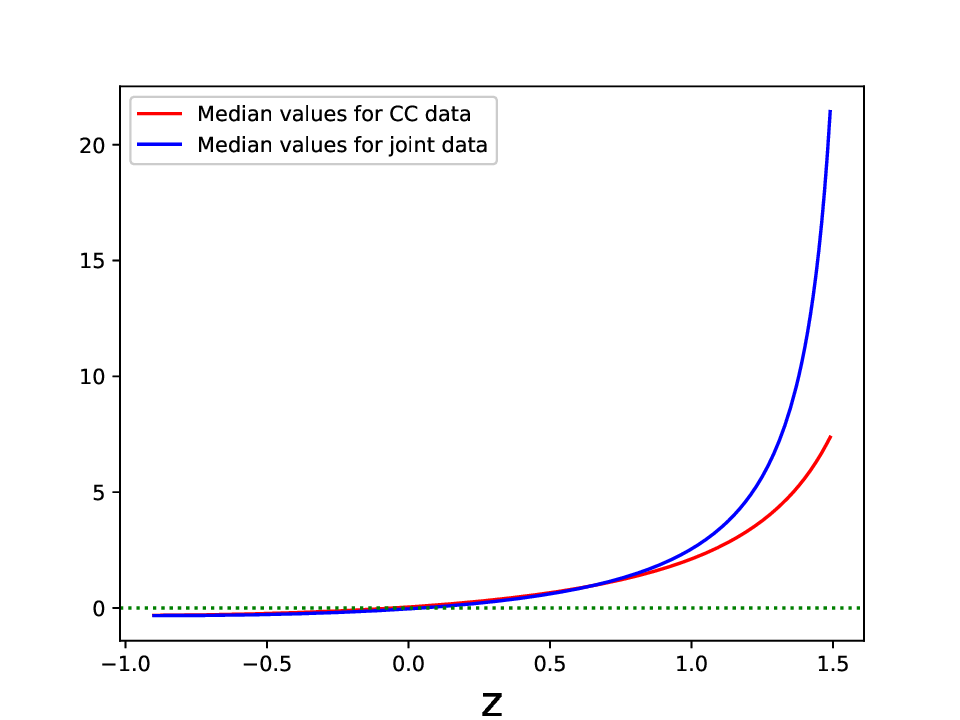}
		\caption{For Model(I): $(\rho + 3p)$ versus $\mathit{z}$.}
		\label{fig:12}
	\end{minipage}\hfill
	\begin{minipage}{0.50\textwidth}
		\centering
		\includegraphics[width=8.2cm,height=7cm]{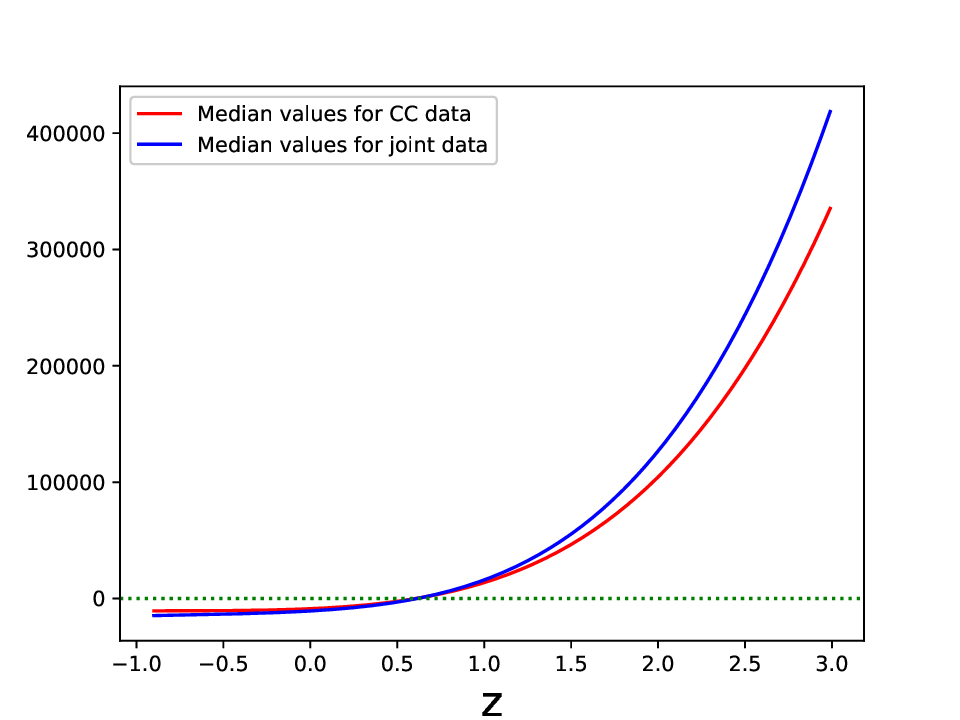}
		\caption{ For Model(II): $(\rho + 3p)$ versus $\mathit{z}$.}
		\label{fig:13}
	\end{minipage}
\end{figure}

\begin{figure}[!htb]
	\captionsetup{skip=0.4\baselineskip,size=footnotesize}
	\begin{minipage}{0.50\textwidth}
		\centering
		\includegraphics[width=8.2cm,height=7cm]{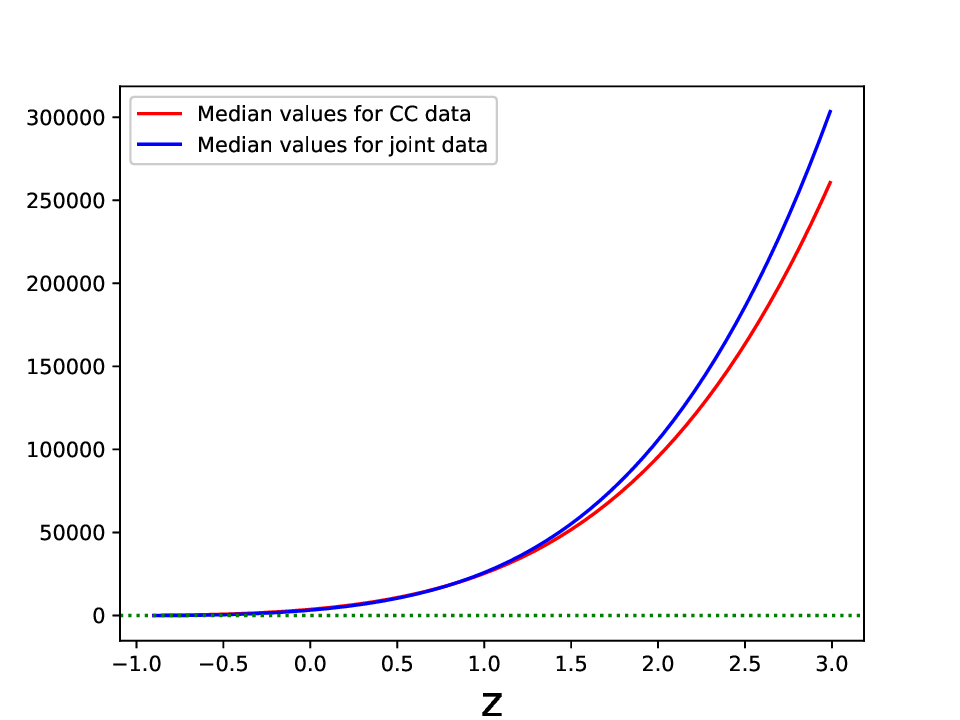}
		\caption{For Model(II): $(\rho + p)$ versus $\mathit{z}$.}
		\label{fig:14}
	\end{minipage}\hfill
	\begin{minipage}{0.50\textwidth}
		\centering
		\includegraphics[width=8.2cm,height=7cm]{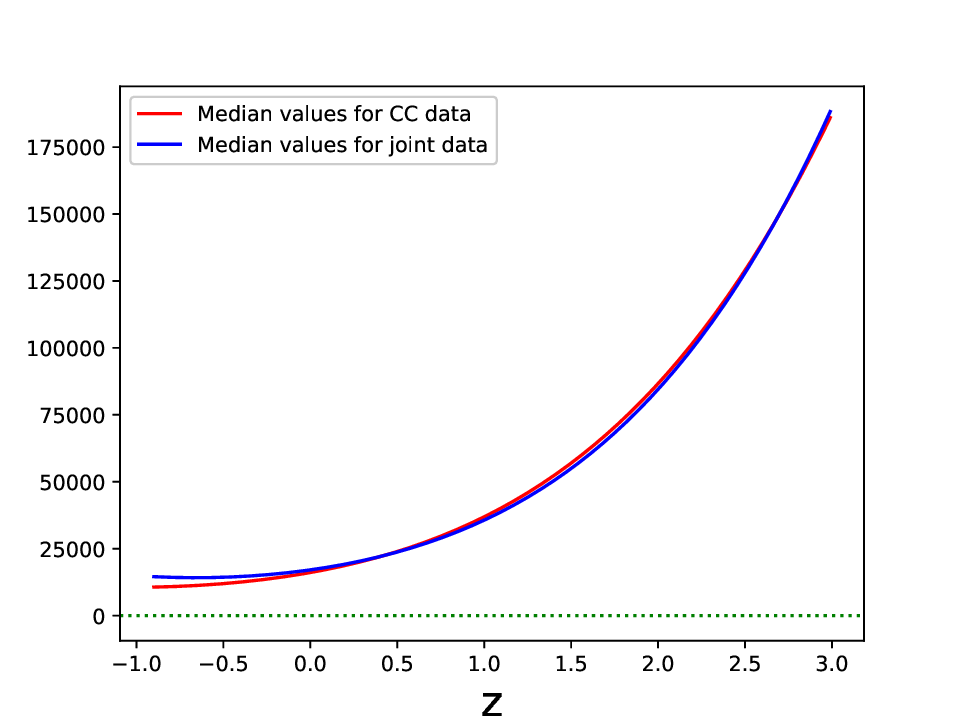}
		\caption{For Model(II): $(\rho - p)$ versus $\mathit{z}$.}
		\label{fig:15}
	\end{minipage}
\end{figure}

In the framework of general relativity and cosmological models, the point-wise energy conditions, based on the stress-energy tensor at any specific spacetime location, impose important constraints on the physical properties of the universe. These energy conditions (EC), essential for ensuring the viability of various models are specified as follows~\cite{visser1997energy,lalke2024cosmic}:
%\vspace{0.2cm}\\
%{\bf{On this specific model's stress-energy tensor, energy conditions represent coordinate-invariant unchanged constraints. The strong energy requirement depends on the Raychaudhuri equation, which also recommends that gravity is attracting ~\cite{singh2022lagrangian}. These requirements inspect the clarity of space-like, null, time-like, and light-like geodesics, that are essential for characterizing the fluid being allocated with~\cite{singh2022cosmological, bouhmadi2015little}. The following are critical parameters meant for the universe's pressure and energy density in a model~\cite{singh2022lagrangian}:}}
%\vspace{.3cm}\\
%\begin{itemize}
% \item[Null:] The requirement that $\rho + p \geq 0$ is known as the null EC (NEC) and signifies that energy and pressure together yield a positive or zero result.
% 
%\item[Weak:] The weak EC (WEC) requires that $\rho \geq 0$ and  $ \rho + p \geq 0$, meaning that neither the energy density nor its sum with pressure can be negative.
%
%\item[Strong:] The Strong EC (SEC) implies that the combination of energy density and pressure must satisfy two inequalities: $\rho + 3p \geq 0$ and $\rho + p \geq 0$, ensuring that the gravitational influence of matter remains attractive.
%
%\item[Dominant:] Dominant EC (DEC) specifies that the energy density($\rho$) is non-negative and must be greater than or equal to the absolute value of pressure, i.e. $ \rho  \geq |p| $.
%
%\end{itemize}
%%%%%%%%%%%%%%%%%%%%%%%%%%%%%%%%%%%%%%%%%%%%%%%%%%%%%%%%%%%%
\begin{itemize}
	\item \textbf{Null Energy Condition:} The condition $\rho + p \geq 0$ is known as the null EC (NEC) and signifies that the sum of energy density and pressure remains non-negative.
	
	\item \textbf{Weak Energy Condition:} The weak EC (WEC) states that $\rho \geq 0$ and  $ \rho + p \geq 0$, meaning that neither the energy density nor its sum with pressure can be negative.
	
	\item \textbf{Dominant Energy Condition:} Dominant EC (DEC) specifies that the energy density($\rho$) is non-negative and must be greater than or equal to the absolute value of pressure, i.e. $ \rho  \geq |p| $.
	
	\item \textbf{Strong Energy Condition:} The strong EC (SEC) implies that the combination of energy density and pressure must satisfy two inequalities: $\rho + 3p \geq 0$ and $\rho + p \geq 0$, ensuring that the gravitational influence of matter remains attractive.
	
\end{itemize}
%%%%%%%%%%%%%%%%%%%%%%%%%%%%%%%%%%%%%%%%%%%%%%%%%%%%%%%%%%%%
In the context of the Universe's decelerated expansion phase, the effective gravitational mass term $(\rho+3p)$ is expected to maintain a positive value. Nonetheless, a growing body of observational data suggests that this strong EC might have been violated during the transitional epoch between galaxy formation and the current cosmological era. Such a deviation implies the likely emergence of negative pressure components within galactic structures, potentially manifesting as anti-gravitational phenomena that contribute to the Universe's accelerated expansion. 
%{\bf{Accordingly, the universe's evolution in an accelerating proportion and the related fact of repulsive gravity more likely to be clarified by the condition $ \rho + 3p < 0 $.}} 
%\vspace{0.3cm}\\
 Given that the SEC consists of two interdependent inequalities, a violation of any one directly results in the failure of the Strong EC \cite{singh2022lagrangian,singh2023homogeneous}. Energy condition frameworks specific to Model(I): $f(R, L_{m}) = \frac{R}{2}+(1+\eta R) L_{m}$ and Model(II): $f(R, L_{m}) = \frac{R}{2}+ L_{m}^{\eta}$ are graphically illustrated in figures (\ref{fig:10}) to (\ref{fig:12}) and (\ref{fig:13}) to (\ref{fig:15}), respectively. Furthermore, the quantity $ (\rho + 3p) $ shifts from positive to negative values at redshifts $ z = -0.04  $ for the CC dataset and $ z = 0.05 $ for the joint dataset in model(I). Similarly, this transition occurs at $ z = 0.6073 $ (For CC dataset) and $ z=0.614 $ (For joint dataset) in model(II). These transition points indicate that both models are consistent with the fulfillment of the fundamental energy conditions (EC), namely the Null, Weak, Dominant and Strong EC, within their respective cosmological evolution.
%Furthermore, the trajectory of $(\rho +3p)$ shifts from positive to negative values around $z=0.6607$. 
%\vspace{0.2cm}\\
%{\bf{In together the present-day and future setups, $ (\rho + p) $ and $ (\rho + 3p) $ draw the negative values, whereas $ (\rho - p) $ draws positive values. The violation of the NEC(Null Energy Condition)  proposes the existence of dark material. For ordinary matters, NEC always holds the properties. The exotic matter that defines wormholes in classical GR is illustrious by a stress-energy tensor that interrupts the NEC. \cite{morris1988wormholes,visser2000energy}.}} 
%\vspace{0.2cm}\\
%{\bf{The validity of the NEC specifies the prospect of either decelerating or accelerating expansion owing to quintessence-like dark energy. For all the inhibited values obtained from the CC datasets, the condition $ (\rho + P) \geq 0 $ holds. Though, for the constrained values from the joint dataset, the violation of $ (\rho + P) \geq 0 $ recommends the existence of phantom-like dark energy in this model. Briefly, we can say that the phantom kind of dark energy might be omitted depending on the violation of $ (\rho + P) \geq 0 $ for the model.}}

%%%%%%%%%%%%%%%%%%%%%%%%%%%%%%%%%%%%%%%%%%%%%%%%%%%%%%%%%%%
%%%%%%%%%%%%%%%%%%%%%%%%%%%%%%%%%%%%%%%%%%%%%%%%%%%%%%%%%%%
%%%%%%%%%%%%%%%%%%%%%%%%%%%%%%%%%%%%%%%%%%%%%%%%%%%%%%%%%%%
%\begin{figure}
%\includegraphics[width=11.5cm,height=5.5cm]{KHOMESHSECFORMODEL2.eps}
%\caption{Model(I): Variation of $(\rho + 3p)$ with redshift($\mathit{z}$)  }
%\label{fig:13}
%\end{figure}
%%%%%%%%%%%%%%%%%%%%%%%%%%%%%%%%%%%%%%%%%%%%%%%%%%%%%%%%%%%
%%%%%%%%%%%%%%%%%%%%%%%%%%%%%%%%%%%%%%%%%%%%%%%%%%%%%%%%%%%
%\begin{figure}
%	\includegraphics[width=11.5cm,height=5.5cm]{KHOMESHSECFORMODEL1.eps}
%	\caption{Model(II): Variation of $(\rho + 3p)$ with redshift($\mathit{z}$)  }
%	\label{fig:16}
%\end{figure}

\newpage

\subsection{Cosmographic parameters}\label{sec:6.5}
\begin{figure}[!htb]
	\captionsetup{skip=0.4\baselineskip,size=footnotesize}
	\begin{minipage}{0.50\textwidth}
		\centering
		\includegraphics[width=8.2cm,height=7cm]{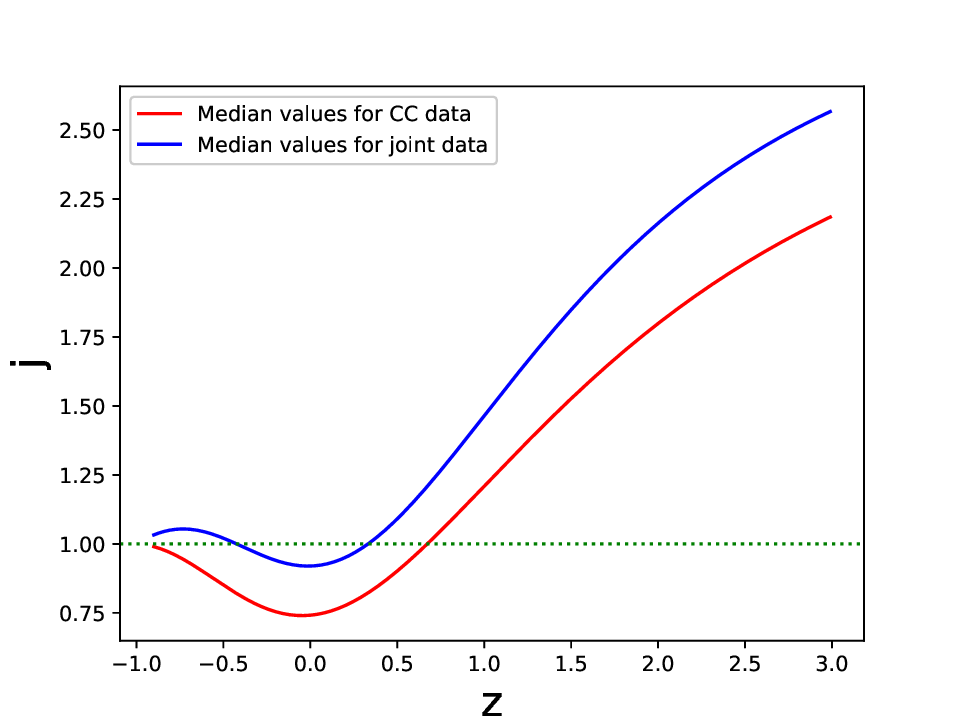}
		\caption{Variation of jerk parameter ($\mathit{j}$) versus $\mathit{z}$.}
		\label{fig:16}
	\end{minipage}\hfill
	\begin{minipage}{0.50\textwidth}
		\centering
		\includegraphics[width=8.2cm,height=7cm]{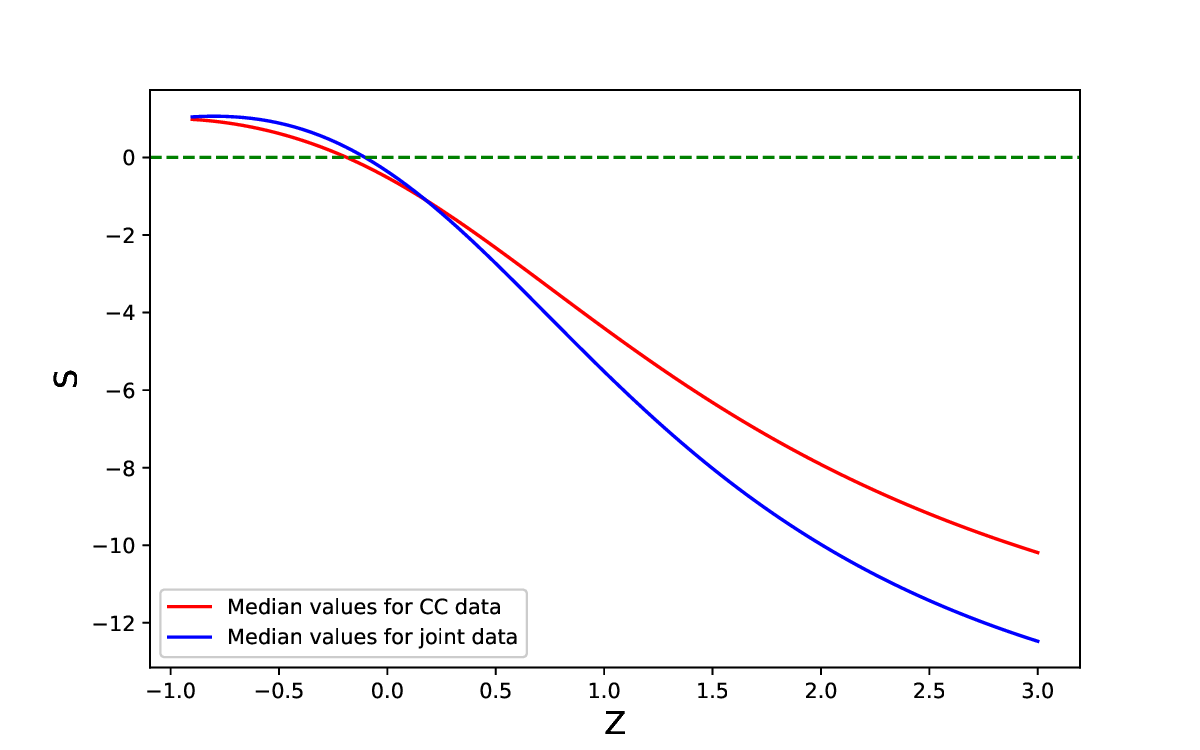}
		\caption{Variation of snap parameter ($\mathit{s}$) versus $\mathit{z}$.}
		\label{fig:17}
	\end{minipage}
\end{figure}

The evolution of the universe can be characterized through kinematic parameters like jerk $\mathit{(j)}$ and snap $\mathit{(s)}$, which are derived from the scale factor and its successive time derivatives, offering a purely geometric perspective on cosmic dynamics. Weinberg~\cite{weinberg2008cosmology} was among the first to propose cosmography as a method to study the universe’s expansion by expressing the scale factor as a Taylor series around the current epoch $t_{0}$, providing a model-independent approach to understanding cosmic evolution.
\vspace{0.2cm}\\
Before evidence emerged for the acceleration of cosmic expansion, the Hubble parameter $(H)$ was viewed as a time-dependent quantity, indicative of the universe's varying expansion rate. The deceleration parameter q offers insight into the changing rate of the universe's expansion, as it is derived from the second time derivative of the scale factor~\cite{mukherjee2016parametric}. This study examines the role of snap $\mathit{(s)}$ and jerk $\mathit{(j)}$ parameters in understanding cosmic evolution. By analyzing jerk also known as jolt and snap alternatively referred to as jounce, we gain valuable insights into the universe's dynamics~\cite{visser2004jerk}:
\begin{equation}{\label{37}}
	\mathit{j}=\frac{1}{aH^{3}}\left(\frac{d^{3}a}{dt^{3}}\right),\ \ \mathit{s}=\frac{1}{aH^{4}}\left(\frac{d^{4}a}{dt^{4}}\right).
\end{equation}
\vspace{.1cm}\\
%In scenarios without spatial curvature, expression (\ref{37}) effectively captures cosmographic analysis.
% Yet, the inclusion of spatial curvature in the model restricts the standard cosmographic methodology. Recent advancements in Pade polynomials~\cite{baker1996pade, gruber2014cosmographic, aviles2014precision, capozziello2022model, capozziello2019model} and Chebyshev polynomials~\cite{capozziello2019extended, capozziello2018cosmographic} offer potential improvements to standard cosmography. Further details are available in~\cite{capozziello2019extended, capozziello2019kinematic, capozziello2020high, bajardi2023late}.
Equation (\ref{37}) is used to describe jerk and snap parameters as functions of redshift
\cite{wang2009probing}:
\begin{equation}{\label{38}}
	j(z)=q(z)\left(1+2q(z)\right)+\frac{dq}{dz}(1+z), \  \ \  \   s(z)=-j(z)\left(2+3q(z)\right)-\frac{dj}{dz}(1+z).
\end{equation}
Figures (\ref{fig:16}) and (\ref{fig:17}) depict the evolution of the jerk and snap parameters respectively, based on the median values of the model parameters. In the current analysis, the median values of the jerk and snap parameters derived from the CC dataset are $\mathit{j_{0}}= 0.7416 $ and $\mathit{s_{0}}=-0.5175 $ respectively. Notably, the jerk parameter value deviates from the $\Lambda$CDM model's expected value of $\mathit{j_{0}}= 1 $, indicating potential differences in cosmic acceleration behavior. Furthermore, when considering the joint dataset, the median values are $\mathit{j_{0}}= 0.9197 $ and $\mathit{s_{0}}=-0.3616 $, suggesting a closer alignment with the $\Lambda$CDM model. According to the obtained results, the trends indicate that this model deviates from the $\Lambda$CDM model in the early universe but approaches it in the later stages of cosmic evolution.

%%%%%%%%%%%%%%%%%%%%%%%%%%%%%%%%%%%%%%%%%%%%%%%%%%%%%%%%%%%
%\begin{figure}[h!]
%    \centering
%    \includegraphics[width=0.8\textwidth]{KHOMESHJERK.eps}
%    \caption{Variation of jerk parameter (j) with redshift(z) }
%    \label{fig:17}
%\end{figure}
%%%%%%%%%%%%%%%%%%%%%%%%%%%%%%%%%%%%%%%%%%%%%%%%%%%%%%%%%%%

%%%%%%%%%%%%%%%%%%%%%%%%%%%%%%%%%%%%%%%%%%%%%%%%%%%%%%%%%%%%%%%
\subsection{Age of the Universe}\label{sec:6.6}
The evolution of the cosmic age $t(z)$ as a function of redshift $ \mathit{z} $ is expressed in \cite{tong2009cosmic} for the cosmological model.
\begin{equation} {\label{39}}
t(z) = \int_{z}^{\infty} \frac{dz_{1}}{(1+z_{1})H(z_{1})}.
\end{equation}
Employing the Hubble function $ H(z) $ from Eq. (\ref{18}) at the present epoch ($ z = 0 $), we compute the integral to determine the current cosmic age. According to this model, the estimated age of the universe is  $t_{0} =12.97^{+1.63}_{-1.25}$ Gyr based on the CC dataset and $t_{0} =12.81^{+1.51}_{-1.17}$ Gyr using the joint (CC+Pantheon) dataset.
%%%%%%%%%%%%%%%%%%%%%%%%%%%%%%%%%%%%%%%%%%%%%%%%%%%%%%%%%%%%%%%%%
\section{Conclusions}\label{sec:7}
%At present, the observed accelerated expansion of the universe continues to be a central focus in cosmological research, with extensive investigations into various dynamical dark energy models and alternative gravity theories. 
In this work, we concentrate on spatially flat, homogeneous and isotropic FLRW cosmologies, incorporating a perfect fluid within the context of $ f(R, L_{m})$  gravity. To examine the effects of this theory on cosmic evolution, we explore two specific non-linear functional forms: $f(R, L_{m}) = \frac{R}{2}+(1+\eta R) L_{m}$ (Model(I)) and $f(R, L_{m}) = \frac{R}{2}+ L_{m}^{\eta}$ (Model(II)), where $\eta$ denotes a free parameter. This study examines the late-time accelerated expansion of the universe by adopting a parametric expression for the deceleration parameter: $q(z)= -1 + \frac{(1+z)(a+2bz)}{(1+az+bz^{2})}$, where $a$ and $b$ are model parameters. From this parametrization, we obtain a quadratic form of the normalized Hubble parameter as $H(z)= H_{0}(1+az+bz^{2})$. This functional form is subsequently embedded into the modified Friedmann equations within the $ f(R, L_{m})$  gravity framework to explore its implications for the dynamics of cosmic expansion. In addition, the deceleration parameter was analyzed to study the universe’s expansion behavior. The findings indicate a recent shift from a decelerating phase to an accelerating phase, supporting the evidence for late-time cosmic acceleration. The present-day deceleration parameter is estimated as $q_{0}=-0.4599$ (for the CC data set) and $q_{0}=-0.542$ (for the joint data set) and the value of redshift at which the transition occurs is $z_{t}=0.655$ (for both the CC and joint data sets). 
\vspace{0.2cm}\\
Moreover, the median estimates of the model parameters were derived for both observational data sets. The obtained values are $H_{0}=67.8^{+1.7}_{-1.7} Km/(s.Mpc)$, $a=0.54^{+0.11}_{-0.11}$, $b=0.265^{+0.058}_{-0.060}$ (for CC data set) and $ H_{0}=68.7^{+1.9}_{-1.9} Km/(s.Mpc)$, $a=-0.458^{+0.061}_{-0.061}, $  $b=0.313^{+0.080}_{-0.080}$ (for joint data set). 
%able with p  for the joint data sets correspondingly.
%Furthermore, we obtained the model parameter's best-fit values using the CC data set and joint data. The obtained best fit values are   : $h_{0}= 66.6^{+1.1}_{-1.1} \ Km/(s.mpc)$\cite{solanki2023cosmic}, \ $ w_{0}= -1.24^{+0.17}_{-0.17}$~\cite{hinshaw2013nine},\  $w_{1}= 0.338^{+0.068}_{-0.075}$~\cite{hinshaw2013nine}, \  $n= 1.03^{+0.17}_{-0.062}$~\cite{jaybhaye2022cosmology}{\bf{(for CC data)}} and $h_{0}= 68.7^{+1.9}_{-1.9} \ Km/(s.mpc)$~\cite{lalke2024cosmic},  $ w_{0}= -1.64^{+0.24}_{-0.24}$~\cite{hinshaw2013nine}, \   $w_{1}= 0.48^{+0.11}_{-0.11}$~\cite{hinshaw2013nine}, \  $n= 1.02^{+0.17}_{-0.079}$~\cite{jaybhaye2022cosmology} {\bf{(for joint data set)}}.  
\vspace{0.2cm}\\
Both models exhibit a physically viable behavior, with the energy density remaining positive, while the pressure likely became negative in recent times. This characteristic aligns with the expected dynamics of late-time cosmic acceleration, indicating that the models effectively capture the essential features of the present-day accelerating universe. In the present epoch, the value of the equation of state parameter for Model(I) yields $\omega_{0}=-0.2800$ (CC data set) and $\omega_{0}=-0.3893$ (joint data set), while for Model(II) gives $\omega_{0}=-0.6295$ (CC data set) and $\omega_{0}=-0.6857$ (joint data set), respectively. These results indicate that Model(II) provides stronger late-time acceleration and better agreement with dark energy dominance as compared to Model(I). Model(II) currently exhibits quintessence-like behavior.    
%\vspace{0.2cm}\\
The energy conditions are checked to understand the stability of the models. All conditions, except the SEC, show positive results, while the SEC is violated. This violation aligns with the observed change in the universe's expansion from deceleration to acceleration. Nonetheless, both models match the current observations of the universe’s acceleration.
%We have discussed the characteristics of physical parameters, i.e., ($\rho$ and $p$),
\vspace{0.2cm}\\ 
In the asymptotic future $(z \to -1)$, the $\Lambda$CDM model yields the cosmological parameters $q=-1$, $\omega=-1$, $j=1$, and snap(s)$=1$, reflecting a de Sitter-like behavior. Similarly, the de Sitter model, characterized by a constant Hubble parameter, also results in $q=-1$, $j=1$, and $s=1$, indicating a consistent late-time cosmic expansion.
Ultimately, the median values of the jerk and snap parameters from the CC dataset are $\mathit{j_{0}}= 0.7416 $ and $\mathit{s_{0}}=-0.5175 $ respectively, indicating a deviation from the $\Lambda$CDM prediction of $\mathit{j_{0}}= 1 $. For the joint dataset, the values $\mathit{j_{0}}= 0.9197 $ and $\mathit{s_{0}}=-0.3616 $ show closer agreement with $\Lambda$CDM model. The results suggest that the model departs from the $\Lambda$CDM scenario at early times but converges toward it in the late universe. Based on the adopted model, the present age of the universe is computed as $t_{0} =12.97^{+1.63}_{-1.25}$ Gyr and $t_{0} =12.81^{+1.51}_{-1.17}$ Gyr for the CC and joint datasets, respectively. Overall, the chosen parametric form of the deceleration parameter within the $ f(R, L_{m})$  gravity framework provides a viable and compelling approach to account for the observed late-time acceleration of the universe.
\section*{\textbf{Acknowledgements}}
One of the authers, G. P. Singh gratefully acknowledges the support provided by the Inter-University Centre for Astronomy and Astrophysics (IUCAA), Pune, India, under the Visiting Associateship Programme.

%R. Garg thanks to Dr. Ashutosh Singh for the insightful discussions. 
%R. Garg would like to thank the Inter-University Centre for Astronomy and Astrophysics(IUCAA), Pune India for providing the IUCAA's Visiting Program under which a part of this work was carried out.

%%%%%%%%%%%%%%%%%%%%%%%%%%%%%%%%%%%%%%%%%%%%%%%%%
\bibliographystyle{unsrt}
\bibliography{PAPER_WRITING}

\end{document}